\documentclass{iopart}
\usepackage{iopams}
\usepackage[dvips]{graphicx}
\usepackage{amstext}
\def\ten#1{\mbox{\boldmath$#1$}}
\usepackage{epsfig}

\begin{document}

\title[Interplay of inequivalent atomic positions in resonant x-ray diffraction]{Interplay of inequivalent atomic positions in resonant x-ray diffraction of Fe$_3$BO$_6$}
\author{G. Beutier$^1$, E. Ovchinnikova$^2$, S. P. Collins$^1$, V. E. Dmitrienko$^3$,
J. E. Lorenzo$^4$, J.-L. Hodeau$^4$, A. Kirfel$^5$, Y. Joly$^4$, A. A. Antonenko$^2$, V. A. Sarkisyan$^{3,6}$, A. Bombardi$^1$}
\address{$^1$ Diamond Light Source, Harwell Science \& Innovation Campus, OX11 0DE, United Kingdom.}
\address{$^2$ Department of Physics of Moscow State University, 119899 Moscow, Russia}
\address{$^3$ A.V. Shubnikov Institute of Crystallography, 119333, Moscow, Russia}
\address{$^4$ Institut Louis N\'eel, CNRS, BP 166X, F-38042 Grenoble, France}
\address{$^5$ Steinmann Institute, University of Bonn, Poppelsdorfer Schloss, D-53115 Bonn, Germany}
\address{$^6$ now at Beam Engineering for Advanced Measurements Co., 809 South Orlando Ave., Suite I, Winter Park, FL 32789, USA.}
\ead{guillaume.beutier@diamond.ac.uk}

\begin{abstract}

"Forbidden" Bragg reflections of iron orthoborate Fe$_3$BO$_6$
were studied theoretically and experimentally in the vicinity the iron $K$-edge.
Their energy spectra are explained as resulting from the
interference of X-rays scattered at two inequivalent crystallographic sites occupied by iron ions.
This particular structure property gives rise to complex azimuthal dependences of the reflection intensities in the pre-edge region as they result from the interplay of site specific dipole-quadrupole and quadrupole-quadrupole resonant scattering.
Also evidenced is an anisotropic character of the absorption spectrum. Self-absorption correction to the diffraction data, as well as possible
contributions of thermal vibrations and magnetic order, are discussed.
Particular care is given to extracting clean spectra from the data, and it
is demonstrated that excellent results can be obtained even from measurements that appear corrupted by several effects such as poor crystal quality and multiple scattering.

\end{abstract}


\submitto{\JPCM}

\maketitle

\section{Introduction}

Resonant X-ray Scattering (RXS) has emerged in the last two decades
as a popular method for studying local properties of crystals,
such as magnetic, charge and orbital ordering, thermal vibrations etc.
\cite{hodeau,lovesey05,review05,collins07}.
The anisotropy of the tensor of scattering in the vicinity of an absorption edge
allows for the observation of Bragg reflections that are forbidden in the isotropic
case by glide-plane and/or screw-axis symmetries.
Such reflections were first observed by D.H. and L.K. Templeton \cite{templeton85a}
and theoretically described in \cite{dmitrienko83,dmitrienko84}.
Since then, "forbidden" reflections were studied experimentally in many crystals.
They are also referred to as pure resonant
reflections, because they appear only very close to absorption edges
owing to resonant scattering processes.
Because of this energy constraint and of their weakness, they can be measured only with synchrotron radiation.

The sensitivity of RXS to local anisotropy arises from
multipole electronic transitions and is hence described in terms of scattering tensors of various ranks.
The strongest contributions to pure resonant reflections are provided by
the so called dipole-dipole scattering.
Beyond this, there are numerous physical phenomena contributing to
the resonant reflections. Higher-order contributions, like
dipole-quadrupole \cite{templeton94,dmitrienko01,dimatteo} or
quadruple-quadrupole \cite{finkelstein,carra} become important when
the dipole-dipole contribution vanishes because of symmetry
restrictions or physical selection rules. In magnetic crystals,
the magnetic reflections are mainly provided by the antisymmetric
part of the dipole-dipole contribution
\cite{gibbs,hannon,namikawa}. The co-existence of both a magnetic
structure and local crystal anisotropy can, in principle,
give rise to so called "combined" forbidden reflections \cite{ovchinnikova00}, which have not been observed so far.
The study of RXS in GdB$_4$ in \cite{ji} revealed both
time-even and time-odd (magnetic) dipole-dipole contributions to forbidden reflections dominating at different temperatures.
The situation becomes even more complicated by the possibility of further effects contributing to forbidden reflections.
Contributions associated with thermal vibrations and point defects were predicted in \cite{dmitrienko99,dmitrienko00},
and thermal-motion-induced (TMI) reflections were observed in Ge \cite{kokubun2001,kirfel02,detlefs}, ZnO \cite{collinstmi} and GaN \cite{collinstmi2}.
In addition, there can be non-resonant scattering arising from magnetic modulations.
The interference between non-resonant magnetic scattering and resonant
dipole-quadrupole and quadrupole-quadrupole channels was found
in hematite ($\alpha$-Fe$_2$O$_3$) \cite{kokubun2008}.
Thus, in many cases, the energy spectra of forbidden reflections can be
explained as the result of the interference between two or more
scattering channels revealing this way an interesting physics of
electronic interactions.
Interference phenomena may be even more important when the
resonant atoms occupy different crystallographic sites.
Such situation was observed in magnetite Fe$_3$O$_4$ \cite{garcia}.
There, the $16(b)$ position of iron provides the $h00, h=4n+2$
reflections corresponding to the dipole-dipole scattering with
a maximum close to  the absorption edge, while the ions in the
$8(a)$ positions allow only for an essential dipole-quadrupole term
in the pre-edge. Thus the peaks corresponding to the different
positions are separated in energy.
For garnets, it was shown that different atomic positions can provide different forbidden reflections \cite{kolchinskaya}.

So far, in all studied cases the interference between various scattering processes or different atomic positions could be disentangled by taking advantage of their manifestation at different energies or thanks to crystallographic selection rules. Here we report the case of iron orthoborate (Fe$_3$BO$_6$), in which scattering from two inequivalent crystallographic positions interferes, and this following the same selection rules and featuring the same tensor components.
Using theoretical analysis and careful data treatment, we demonstrate that RXS is now a technique sufficiently mature to study such a complicated case, even in the presence of experimental complications such as poor crystal quality, self-absorption and multiple-scattering.

\section{Theoretical}

In kinematic diffraction theory, which is  usually applied
to RXS,  a forbidden reflection intensity is simply determined by
the resonant structure amplitude $F(\ten H)$, defined by the resonant scatterer's position and by the
Fourier transform of the atomic scattering factor $f(\ten r)$. The
latter can be decomposed into:
\begin{equation}
    f = f^0+if^m+f^\prime+if^{\prime\prime}
\end{equation}
where $f^0$ is the energy independent non-resonant charge scattering factor and $f^m$ accounts for non-resonant magnetic scattering.
The last two energy dependent terms  describe the real and imaginary parts of the RXS and obtain particular importance  at energies close to an absorption edge.

The tensor character and the tensorial properties of atomic scattering factors near
absorption edges were considered in many studies both in a Cartesian
approach \cite{blume,beldmi} and also using spherical tensors
\cite{hannon,carra,brouder,lovesey}. In the present paper we shall
use the Cartesian formalism.
Witin this framework, the resonant scattering can be written in form of a superposition of the interactions in different multipole  orders
\begin{equation}
    f^\prime+if^{\prime\prime} = e^{\prime*}_j e_k
    \left[D_{jk}-\frac{i}{2}(k_m I_{jkm}-k^{\prime}_m I^*_{kjm})+ \frac{1}{4}k^\prime_m k_n Q_{jkmn}\right]
\end{equation}
where the summation over repeated indices is implied.

The vectors $\ten k$, $\ten k^\prime$ and $\ten e$, $\ten e^{\prime}$ are the wavevectors and polarizations, respectively, of the incident and scattered radiations.
Below we shall also use conventional $\ten\sigma$ and $\ten\pi$ polarization
vectors, normal and parallel, respectively, to the scattering plane.
The multipolar expansion was extended up to the electric quadrupolar term,
omitting magnetic multipolar and higher order electric multipolar contributions,
which can be expected extremely weak \cite{brouder}.
Thus, the resonant scattering tensor consists only of the second rank dipole-dipole tensor $\ten D$, the third rank dipole-quadrupole tensor $\ten I$,
and the fourth rank quadrupole-quadrupole tensor $\ten Q$, all of which possessing tensor elements that are sensitive to the incident radiation energy.

The symmetry properties of these tensors, in the Cartesian formalism, are as following:
i) the  dipole-dipole tensor is symmetric ($D_{jk}=D_{kj}$) for nonmagnetic crystals.
For magnetic crystals, the magnetic character manifests itself as an antisymmetric part;
ii) the time-even dipole-quadrupole tensor can have both a symmetric
and an antisymmetric part with respect to permutation of the
polarization indices. It was shown that the symmetric part of the
time-even dipole-quadrupole tensor gives rise to forbidden
reflections in Ge and ZnO
\cite{templeton94,kokubun2001,kirfel02,collinstmi} whereas the
antisymmetric part causes some peculiarities
of the forbidden reflections in Fe$_2$O$_3$ and Cr$_2$O$_3$
\cite{dmitrienko01,dimatteo,kokubun2008}.
A time-odd third rank tensor can also occur in the case of a magnetic crystal;
iii) in non-magnetic crystals, the fourth rank  quadrupole-quadrupole tensor is symmetric, i.e. invariant against the permutation of
polarization or wavevector indices \cite{blume}:
$Q_{jknm}=Q_{kjnm}=Q_{jkmn}=Q_{kjmn}=Q_{nmjk}$.
These symmetry properties reduce the number of "distinct" cartesian components, for a non-magnetic crystal,
to 6, 18 and 21 for the dipole-dipole, dipole-quadrupole, and quadrupole-quadrupole tensors, respectively \cite{sirotin}.
However, spherical formalism shows that the number of independent components are 6, 15 and 15 respectively.
The remaining distinct Cartesian components are therefore linearly related, although their linear relationship is not trivial \cite{stone}.

Generally, the intensity of a forbidden reflection is dominated by the dipole-dipole contribution.
However, when this contribution
vanishes, e.g. due to crystal symmetry, the higher order
terms become the essential ones.
They are also important in the pre-edge region, where the dipole-dipole contribution is weak:
quadrupole transitions are reduced in energy due to multi-electron effects such as screening.

Since the tensorial character of the resonant scattering factor causes the reflection intensities to obtain azimuthal dependencies it is essential to identify the non-vanishing tensor elements in order to understand the azimuthal properties of forbidden reflections and their energy dependencies.
Exactly this analytical task is completed in the following section for the title compound, Fe$_3$BO$_6$.

\section{Pure resonant reflections in iron orthoborate}

The symmetry of iron orthoborate structure is described by the space group $Pnma$ (no 62) \cite{white}.
The Fe  cations occupy two octahedral positions:
a special one, $4(c)$ with point symmetry $m$, the mirror plane $m$ is normal to [010],
and a general one, $8(d)$, with point symmetry 1 (ie. no symmetry restrictions).
The resonant structure factor $F(\ten H)$ thus sums  the contributions of the 4 and 8, respectively, Fe cations of both groups:
\begin{equation}
    F(\ten H)=F^{4(c)}(\ten H)+F^{8(d)}(\ten H).
\end{equation}
Naturally, the ratio between the partial structure factors, $F^{4(c)}$ and $F^{8(d)}$, varies when the scattering factors vary with the reflection indices.
Moreover, their relative phases vary strongly upon crossing an absorption edge. Below, it will be shown that the complex interplay between the scattering at the two Fe positions can produce constructive or destructive interference,
leading to completely different energy spectra for different $\ten H = hkl$.

The  pure resonant reflections discussed in the present paper are similar
to the pure nuclear reflections of M\"ossbauer radiation, but the energy spectra are quite different.
In the case of nuclear scattering the resonant peaks are very sharp
($\sim 10^{-8}$ eV) while for electronic transitions the
typical width of resonance lines is about 1--10 eV.
Iron orthoborate has indeed been extensively studied with
M\"ossbauer radiation, and the interplay between the two iron
positions was evidenced \cite{kovalenko,tolpekin}.
Because the M\"ossbauer nuclear transition in iron is very sensitive to magnetism,
it allows for the determination of the magnetic structure of
iron orthoborate \cite{bayukov_mag,kovalenko_mag}.
In the case of RXS, we are dealing with
$K$-edge electronic transitions of the $E1$ and $E2$ types, which
are only weakly sensitive to magnetism. We will therefore develop our
tensor analysis treating Fe as  non-magnetic. The validity of this approximation will be demonstrated by the experiment.

\begin{figure}[ht]
    \centering
    \includegraphics*[width=0.49\columnwidth]{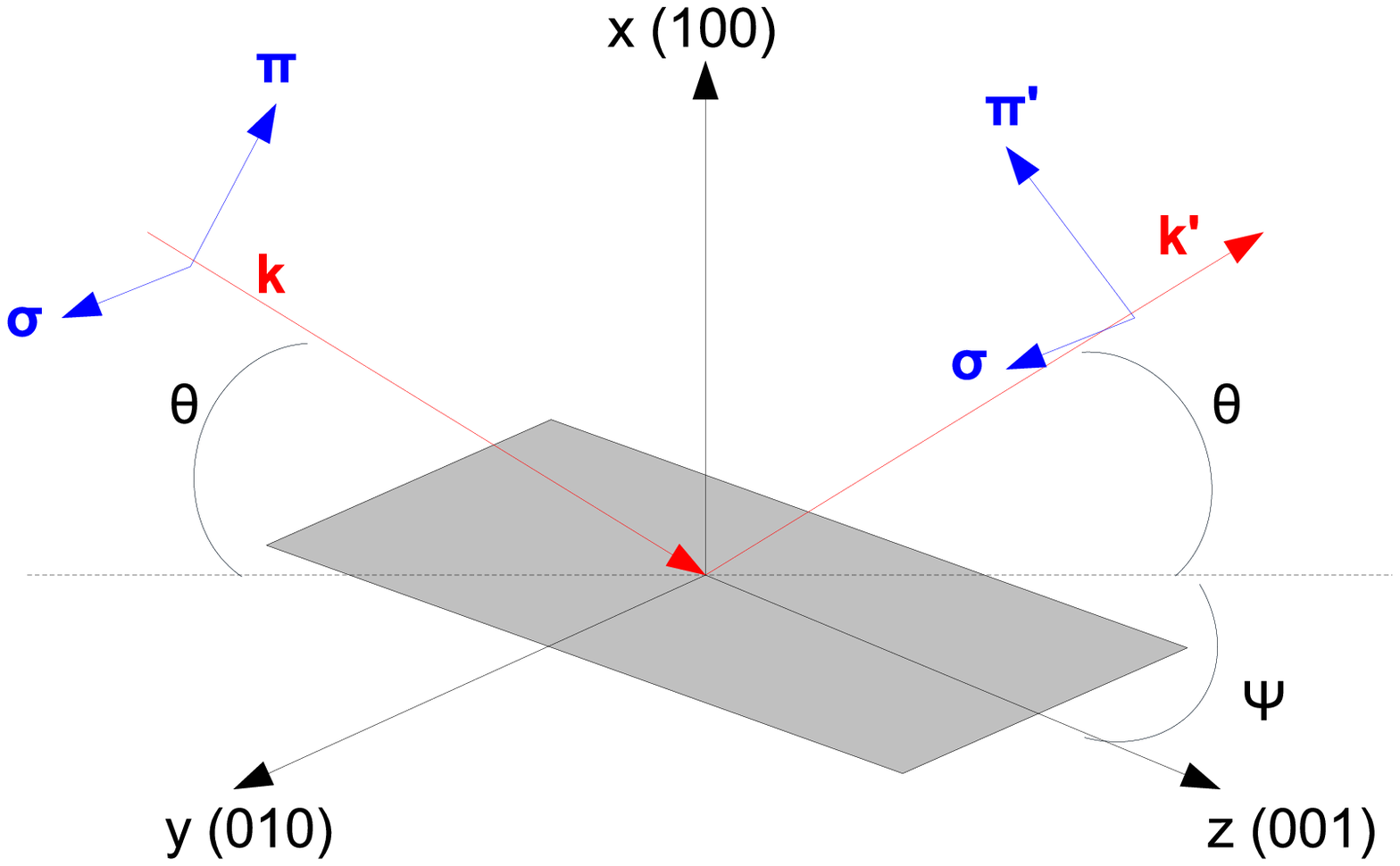}
    \caption{Diagram of the experimental geometry and definition of the coordinate system.
    In this Cartesian system $(x,y,z)$, the vectors $\ten k$, $\ten k^{\prime}$, $\ten \sigma$, $\ten \pi$ and $\ten\pi^{\prime}$ are written as follows:
    \begin{eqnarray}
        && \nonumber \ten k=\mid k\mid (-\sin\theta,-\cos\theta\sin\psi,\cos\theta\cos\psi) \\
        && \nonumber \ten k^{\prime}=\mid k\mid (\sin\theta,-\cos\theta\sin\psi,\cos\theta\cos\psi) \\
        && \nonumber \ten \sigma = (0,\cos\psi,\sin\psi) \nonumber\\
        && \nonumber \ten \pi = (\cos\theta,-\sin\theta\sin\psi,\sin\theta\cos\psi) \\
        && \nonumber \ten \pi^{\prime} = (\cos\theta,\sin\theta\sin\psi,-\sin\theta\cos\psi)
    \end{eqnarray}}
    \label{fig.diagram}
\end{figure}

 Next, we determine the contributions to the resonant
structure factor,  being allowed by  space group symmetry.
We shall first analyse the case of the general position $8(d)$, with general coordinates $(x,y,z)$ in the unit cell , and then proceed to the special position $4(c)$  which second case  can be easily deduced from the first one.
In a general position (i.e.  point symmetry 1), the symmetric $D$, $I$ and $Q$ tensors can have up to 6, 18 and 21, respectively, distinct Cartesian components \cite{sirotin}, some of which vanishing
for particular $hkl$ values.
In the case of $h00,h=2n+1$ reflections, which are the experimentally measured
reflections, only one, $D_{xz}$, of the
six Cartesian components of the dipole-dipole tensor survives and gives rise to
the following contribution to the structure factor,
where the fractional coordinates $(x,y,z)$  refer to  the crystal axes (Fig. \ref{fig.diagram}):
\begin{equation}
    F_{dd}(h=2n+1,0,0) = 8 D_{xz}\cos(2\pi h x)
    \left( \begin{array}{ccc} 0&0&1 \\0&0&0 \\1&0&0\end{array}\right)
    \label{eq.dd}
\end{equation}

Similarly, the dipole-quadrupole tensor has a structure factor with only 5 non-vanishing distinct Cartesian components at the $h00, h=2n+1$ forbidden reflections:
$I_{xxz}$, $I_{yyz}$, $I_{zxx}$, $I_{zyy}$ and $I_{zzz}$.
Describing $\ten k$ and $\ten k^{\prime}$ with respect to the crystal $(x,y,z)$ coordinate system (Fig. \ref{fig.diagram}), we obtain:
\begin{equation}
    F_{dq}(h=2n+1,0,0) = 4\mid \ten k\mid\sin(2\pi h x) \left(\begin{array}{ccc}
    0 & 0 & \tilde I_{xz} \\
    0 & 0 & -\tilde I_{yz} \\
    \tilde I_{xz} & \tilde I_{yz} & 0
    \end{array}\right)
    \label{eq.dq}
\end{equation}
with
\begin{eqnarray}
    && \tilde I_{xz} = -(I_{xxz}+I_{zxx})\sin\theta \\
    && \tilde I_{yz} = (I_{yyz}-I_{zyy})\cos\theta\sin\psi
\end{eqnarray}
The term $I_{zzz}$ vanishes because $k_z=k_z^{\prime}$. For the dipole-quadrupole contribution we see that there are both a symmetric and an antisymmetric part.

Finally, one finds that only 4 of the 21 distinct Cartesian components of the quadrupole-quadrupole tensor contribute to the $h00,h=2n+1$ reflections:
\begin{equation}
    F_{qq}(h=2n+1,0,0) = 2 \mid \ten k \mid^2\cos(2\pi h x) \left(\begin{array}{ccc}
    0 & \tilde Q_{xy} & \tilde Q_{xz} \\
    \tilde Q_{xy} & 0 & 0 \\
    \tilde Q_{xz} & 0 & 0 \\
    \end{array}\right)
    \label{eq.qq}
\end{equation}
with
\begin{eqnarray}
    && \tilde Q_{xy} = -Q_{xyyz}\cos^2\theta\sin2\psi \\
    && \tilde Q_{xz} = -Q_{xxxz}\sin^2\theta +\cos^2\theta\left(Q_{xzyy}\sin^2\psi +Q_{xzzz}\cos^2\psi\right)
\end{eqnarray}

By substituting the polarizations $\sigma=\sigma^\prime$, $\pi$ and $\pi^\prime$ by their Cartesian expressions in equations (\ref{eq.dd}), (\ref{eq.dq}) and (\ref{eq.qq}),
we conclude that the x-rays are scattered only in the rotated channels,
i.e. $\sigma\rightarrow\pi$ and $\pi\rightarrow\sigma$.
For the former one, which is the case of the experimental setup, we obtain the following contributions:
\begin{eqnarray}
    F_{dd}^{\sigma\pi}(h=2n+1,0,0) &=& 8 D_{xz}\cos(2\pi h x)\cos\theta\sin\psi
    \label{eq.ddsp}\\
    F_{dq}^{\sigma\pi}(h=2n+1,0,0) &=& -2 \mid\ten k\mid \left(I_{yyz}-I_{zyy}-I_{xxz}-I_{zxx}\right) \nonumber \\
    && \times \sin(2\pi h x) \sin2\theta\sin\psi
    \label{eq.dqsp}\\
    F_{qq}^{\sigma\pi}(h=2n+1,0,0) &=& 2 \mid\ten k \mid^2 \cos(2\pi h x) \nonumber \\
    && \times \left[Q_1(\theta)\sin\psi + Q_3(\theta) \sin3\psi\right]
    \label{eq.qqsp}
\end{eqnarray}
where $Q_1(\theta)$ and $Q_3(\theta)$ are linear combinations of $Q_{xxxz}$, $Q_{xyyz}$, $Q_{xzzz}$ and $Q_{xzyy}$ with coefficients depending on $\theta$ only.
This analysis shows that the dipole-dipole and dipole-quadrupole amplitudes of the scattered radiation possess a simple $\sin\psi$ azimuthal dependence,
while the quadrupole-quadrupole contribution mixes $\sin\psi$ and $\sin3\psi$ terms.

The above expressions were derived in the case of the general $8(d)$ position, for which no symmetry applies to the atomic site.
We end up with six contributions to the forbidden reflections $h00, h=2n+1$ :
one for the dipole-dipole ($D_{xz}$), one for the dipole-quadrupole (a fixed combination of the $I_{jkn}$),
and four for the quadrupole-quadrupole.
Let us now analyze the case of ions in $4(c)$ positions with point symmetry $m$: they lie on the mirror planes ($y=\frac{1}{4}$ and equivalent planes) which transform $(xyz)$ into $(x,\frac{1}{2}-y,z), etc.$
Thus, all the Cartesian tensor components with indices containing odd powers of $y$ vanish.
However, equations (\ref{eq.ddsp}), (\ref{eq.dqsp}) and (\ref{eq.qqsp}) do not involve such components for  the $h00, h=2n+1$ reflections
so that the scattering description for the cations in the $4(c)$ positions needs  as many components as do  the cations in general $8(d)$ positions.
Consequently, for reflections $h00, h=2n+1$, the resonant amplitude scattered
by each iron site is described by six parameters, i.e. twelve in total, all of which varying with energy.
In principle, it would be possible to determine them all, by measuring all 6 $h00, h=2n+1$ forbidden reflections accessible at the iron $K$-edge energy, at two different azimuths.
But it would be extremely tedious, if experimentally possible, and it is not the purpose of the present article. Here we want to highlight the interference between non-equivalent crystallographic sites.

 The scattering contributions interfere with weights $\cos(2\pi h x)$ for the dipole-dipole and quadrupole-quadrupole, and $\sin(2\pi h x)$ for the dipole-quadrupole, respectively, where
 $x=x_d=0.41246$ for the   iron in the $8(d)$  and $x=x_c=0.12835$ for the other iron  in the $4(c)$ position \cite{diehl}.
The values of the respective coefficients are given in Table \ref{phases}.
Inspecting these values we can see that the dipole-dipole
contribution to the  $300$ reflection is mainly provided by Fe  in  $8(d)$.
 We can further see that the dipole-dipole contributions from   $8(d)$ and $4(c)$ to the $300$ and $500$ reflections have opposite signs while  those to reflection $700$ show comparable values and equal signs.   Thus, the interference between the radiation scattered by the two iron  sites
is of the same type for the reflections  $300$ and $500$,  and
opposite for the $700$ reflection so that significant differences
between the energy spectra of the $300$, $500$ and $700$ reflections can be expected.

\begin{table}
    \centering
    \begin{tabular}{|c|c|c|c|c|}
    \hline $(h,k,l)$& $\cos(2\pi hx_c)$& $\sin (2\pi hx_c)$&$\cos(2\pi hx_d)$&$\sin (2\pi hx_d)$\\
    \hline
    $300$ &  0.0792 & 0.9968&  -0.750 & 0.661\\
    \hline
    $500$ & 0.924 & 0.381 & -0.628 & -0.777\\
    \hline
    $700$ & 0.7591 & -0.650& 0.803 & -0.595\\
    \hline
    \end{tabular}
    \caption{Weights to the contributions in the $h00, h=2n+1$ forbidden reflections of the two sites with positions $x_d=0.12835$ and $x_c=0.41246$. Positions from \cite{diehl}.}
    \label{phases}
\end{table}

\section{Experimental}

The resonant reflections $300$, $500$ and $700$ were measured
at the XMaS beamline of the European Synchrotron Radiation Facility (ESRF) at room temperature. Complementary measurements of the fluorescence yield and of the resonant reflection $700$ at ambient as well as at low and high temperatures were done at the beamline I16 of Diamond Light Source (DLS).
The natural linear polarization of the radiation obtained at the bending magnet (XMaS) or undulator (I16) was used and the energy was tuned to the Fe $K$-edge with a Si(111) monochromator.
At both beamlines, the scattered radiation was measured with silicon-drift detectors, allowing simultaneous records of the elastic scattering and the fluorescence yield in independent channels.
Since, as shown above, the resonant scattering involves a full $\sigma\to\pi$ polarization change it was not necessary to use a secondary polarization analyzer. For all measurements,  the same sample was used:  a platelet with $(\pm 100)$ faces of a few square millimeters size and a thickness of a few hundreds microns.

In the kinematic diffraction theory, which is usually used for the treatment of resonant diffraction,
and in case of a sample larger than the beam and much thicker than the absorption length of the crystal material,
the intensity of the Bragg reflection corresponding to the reciprocal vector $\ten H=hkl$ is equal to
\cite{jeims}:
\begin{equation}
    I(\ten H)\sim\frac {|F(E,\ten e,\ten e',\ten H)|^2}{\mu(E,\ten e)+\mu(E,\ten e')\times g}
    \label{eq.intensity}
\end{equation}
where the linear absorption coefficients, $\mu(E,\ten e)$ and $\mu(E,\ten e')$, depend on both energy $E$ and the polarizations $\ten e,\ten e'$ of the beams,
and $g$ is a geometrical factor defined as:
$g=\frac{\sin\alpha}{\sin\beta}$,
$\alpha$ and $\beta$ are the incident and exit angles with respect to the sample surface.
In the case of $h00$ reflections and a $(100)$ sample surface, $g=1$.
Equation (\ref{eq.intensity}) assumes that the photon polarization is modified only by the scattering process, but not by absorption.
Thus, in order to extract $|F(E,\ten e,\ten e^{\prime},\ten H)|^2$ from the RXS measurements, the measured intensity must be corrected for self-absorption, which requires the determination of the energy dependent absorption and an analysis of its anisotropy.

\subsection{Study of the absorption spectrum}

Fe$_3$BO$_6$ has a crystal structure with point group $mmm$.
The resonant absorption is therefore not isotropic.
The dipolar  component is trichroic (3 independent spectra)
and the quadrupolar one contains 6 independent spectra.
The mixed dipole-quadrupole component on the other hand vanishes because the point group permits only parity-even absorption events.

Let us consider the azimuthal dependence of the absorption spectra about the [100] direction, with azimuthal reference [001], for any incidence angle $\theta$ and any polarization.
Though the polarization vector is generally not an eigenstate of the optical system, we can make this approximation if the anisotropy is not too strong compared to the isotropic absorption.
Within this approximation, the linear absorption coefficient $\mu$ is proportional to the absorption cross-section.
According to Brouder's formalism \cite{brouder}, $\mu$  can be written as:
\begin{eqnarray}
    \mu(E,\theta,\psi) &=& \mu_{nres}+\mu_K(E,\theta,\psi) \\
                       &=& \mu_{nres}+\mu_0(E,\theta)+\mu_2(E,\theta)\cos(2\psi)+\mu_4(E,\theta)\cos(4\psi) \nonumber
\end{eqnarray}
where the non-resonant part of the absorption
$\mu_{nres}$ is almost constant across the $K$-edge, whereas the
resonant part at the $K$-edge, $\mu_K$, is different for different eigenpolarizations.
Since $\mu_0$ and $\mu_2$ are largely dominated by the dipole contribution they can be considered as independent of $\theta$ in case of $\sigma$ polarized light.
The term $\mu_4$ is of quadrupolar origin and occurs only in the pre-edge region; $\mu_2$ and $\mu_4$ are later referred to as 2-fold and 4-fold parts, respectively.

\begin{figure}[ht]
    \includegraphics*[width=0.49\columnwidth]{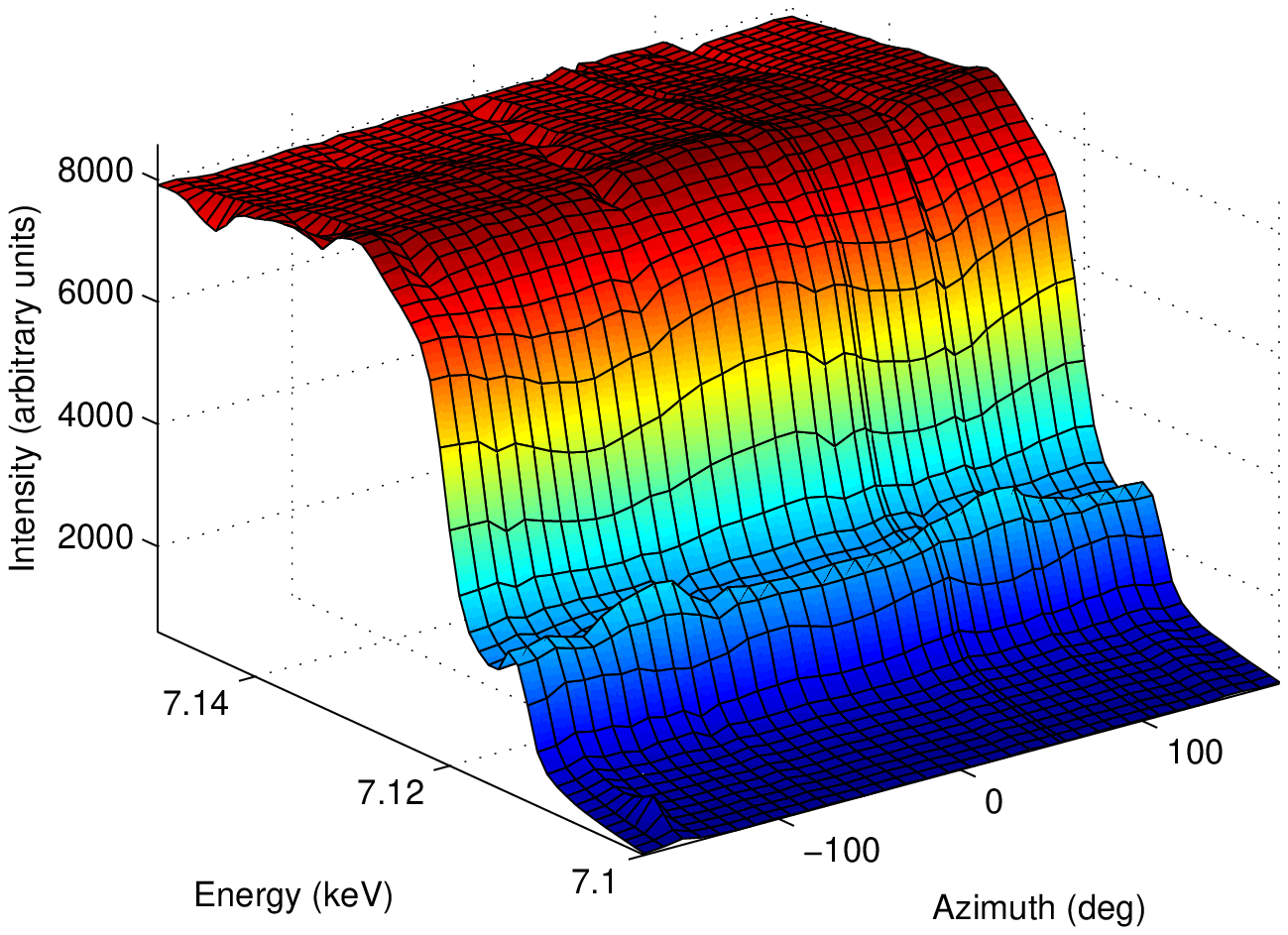}
    \includegraphics*[width=0.49\columnwidth]{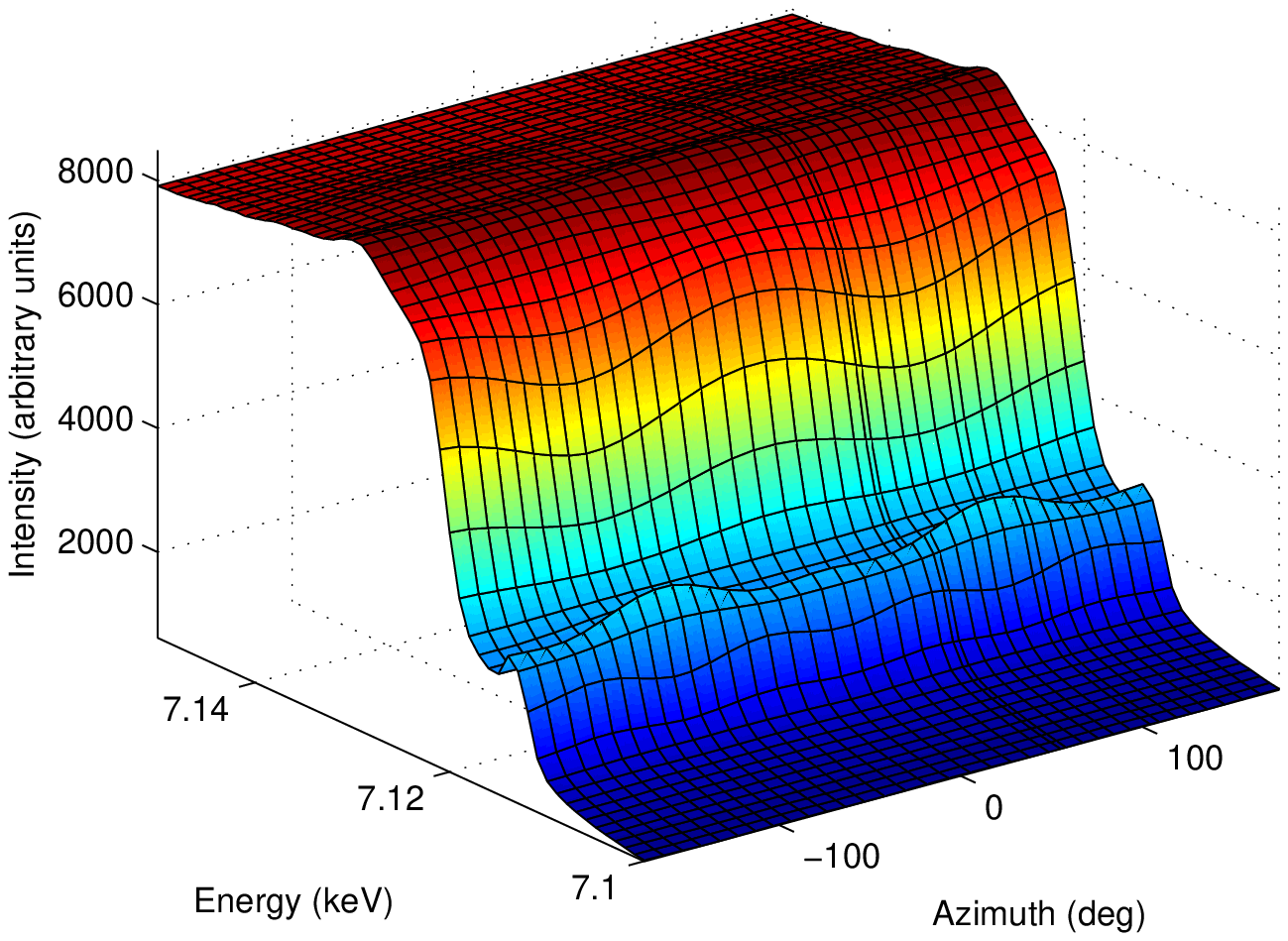}
    \includegraphics*[width=0.49\columnwidth]{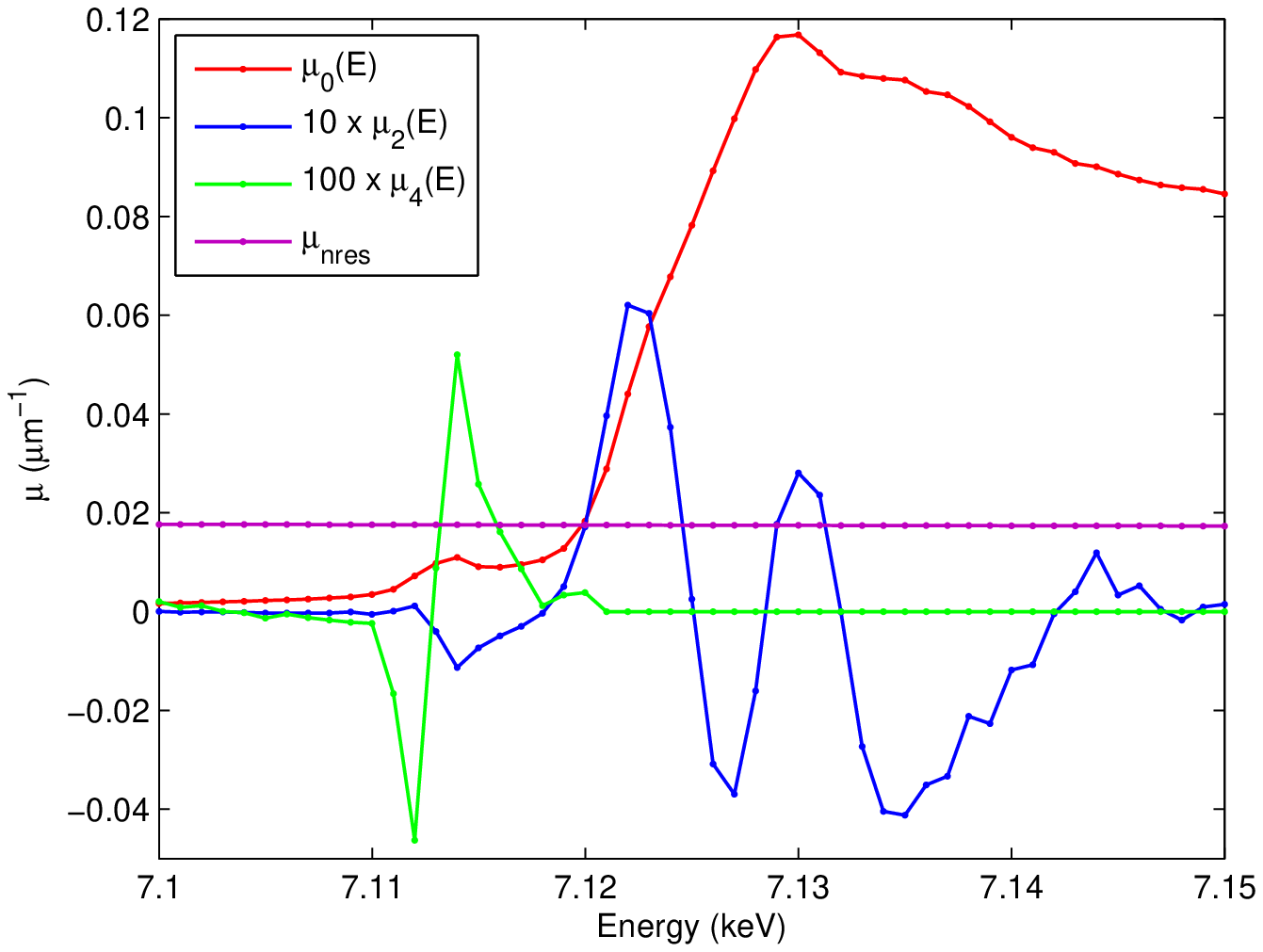}
    \includegraphics*[width=0.49\columnwidth]{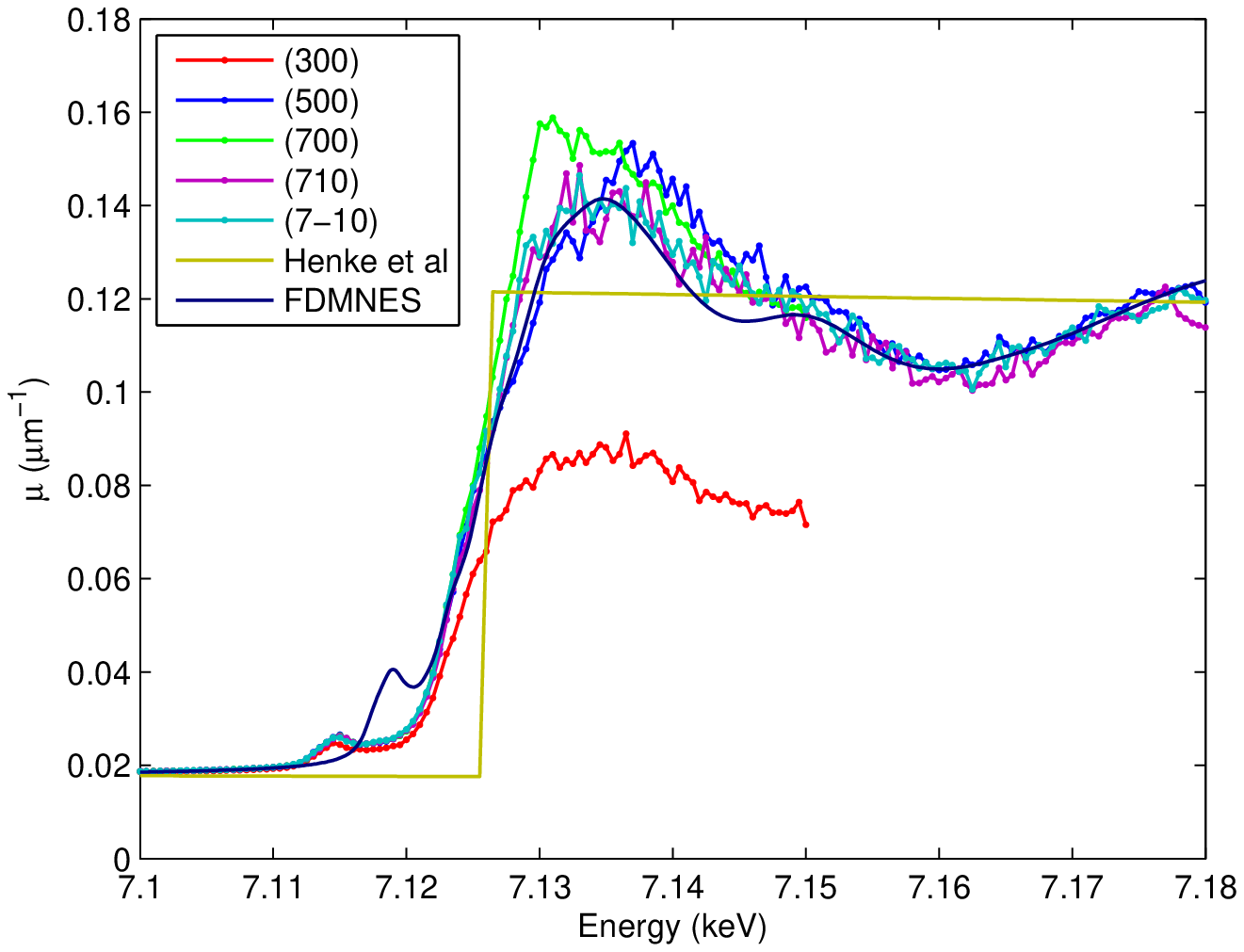}
    \caption{Upper left: Energy-azimuth map of the fluorescence of Fe$_3$BO$_6$ at the Fe $K$-edge and about the $[100]$ direction. The detector is at 90$^\circ$, vertical to the incident beam, the incident angle is around 20$^\circ$ and the polarization is horizontal. Upper right: calculated map from the fit. Lower left: Fit parameters. Lower right: Isotropic part of the absorption spectrum, as obtained from fitting the fluorescence data. The $hkl$ indices refer to the RXS reflections at which the fluorescence data were recorded. Except for reflection 300 the mutual agreement of the fitted spectra is reasonably good. The 300 data set was rather poor due to the combination of the low incidence angle, the large beam size and a sample inhomogeneity. Except for the pre-edge region the general shape of the absorption coefficient is in agreement with the FDMNES calculations (see Section \ref{fit}). For comparison, also the values calculated from Henke's tables \cite{Henke} are  shown.}
    \label{fig.fluo}
\end{figure}

A full energy-azimuth map of fluorescence was measured from the same sample at the DLS beamline I16, with the detector at 90$^\circ$ from the incident beam, the incident angle at 20$^\circ$ and the polarization perpendicular to the $[100]$ axis: for simplicity, we shall now refer to this polarization as $\sigma$, like for the scattering case (Fig. \ref{fig.fluo}, top left).

For a thick sample, the measured fluorescence intensity is \cite{pfalzer}:
\begin{equation}
    \frac{I_f(E)}{I_0(E)} \propto \frac{\mu_K(E)}{\mu(E)+g\mu(E_f)}
\end{equation}
where the angular dependence is implicit. $E_f=6.4$ keV is the average energy of the $K_{\alpha1}$ and $K_{\alpha2}$ fluorescence lines.
The energy-azimuth map was fitted with $\mu_0(E)$, $\mu_2(E)$, $\mu_4(E)$ and a scale factor as free parameters.
$\mu_{nres}(E)$ and $\mu(E_f)$ were calculated from the atomic data tables \cite{Henke}.
The final fit  parameters and the resulting  model map are shown in Fig. \ref{fig.fluo}.
The measured and calculated maps do not show any significative difference, attesting for the good quality of the fit.
The 2-fold and 4-fold parts of the absorption are clearly evidenced and are respectively 1 and 2 orders of magnitude lower than the isotropic part.
Note that these independent spectra are still dependent of $\theta$, and measurements at different incidence angle would be needed to evidence it. However, only the quadrupolar part of the absorption is $\theta$-dependent with this polarization, and it is expected to vanish except around the pre-edge feature. $\mu_0(E)$ and $\mu_2(E)$ are therefore a good approximation of the isotropic and 2-fold components of the dipole absorption for $\sigma$-polarized light.

A similar procedure was applied to the fluorescence data measured along with the RXS data at XMaS: each energy-azimuth map was separately fitted. Since,  however,  the data sets were rather noisy, the 2- and 4-fold components could not be extracted. The isotropic maps are shown in Fig. \ref{fig.fluo}. Although the forbidden reflections change  the light polarization from $\sigma$ to $\pi$, as shown in the theoretical section, the diffracted intensity is so small compared to the incident  intensity that one  can  infer in first approximation that the fluorescence results only from absorption events involving  $\sigma$ polarized photons.  Thus, the  case  described in the previous paragraph applies, and the dipolar absorption should be the same for all $h00$ reflections.

\subsection{RXS data reduction}

For the extraction of $|F(\ten H)|^2$ from the measured data
several effects must be taken into account.

The first one is absorption effect which distorts the energy
spectra. In principle, the absorption has also an angular
dependence, but as shown in the previous section  ignoring it
is not a bad approximation. Hence,  an effective absorption spectrum was
calculated by averaging the 4 reasonably  isotropic absorption
curves derived from the fluorescence measured at the reflections 500, 700, 710 and $7\bar{1}0$ (300 was not considered for the reasons given above). According
to Equation (\ref{eq.intensity}), the measured data were multiplied by
$\mu(E,\ten e)+g \mu(E,\ten e') \approx (1+g)(\mu_{nres}+\mu_0(E))$,
where $g$ is 1 for the $h00$ reflections.
Of course, correcting the data with an
isotropic absorption spectrum and neglecting its azimuthal
dependence introduces errors. From Figure \ref{fig.fluo}, bottom left
panel, we estimate that these can locally amount to  15\% (particularly in
the pre-edge region), and to 5\% at energies above the edge.

\begin{figure}[ht]
    \includegraphics*[width=0.49\columnwidth]{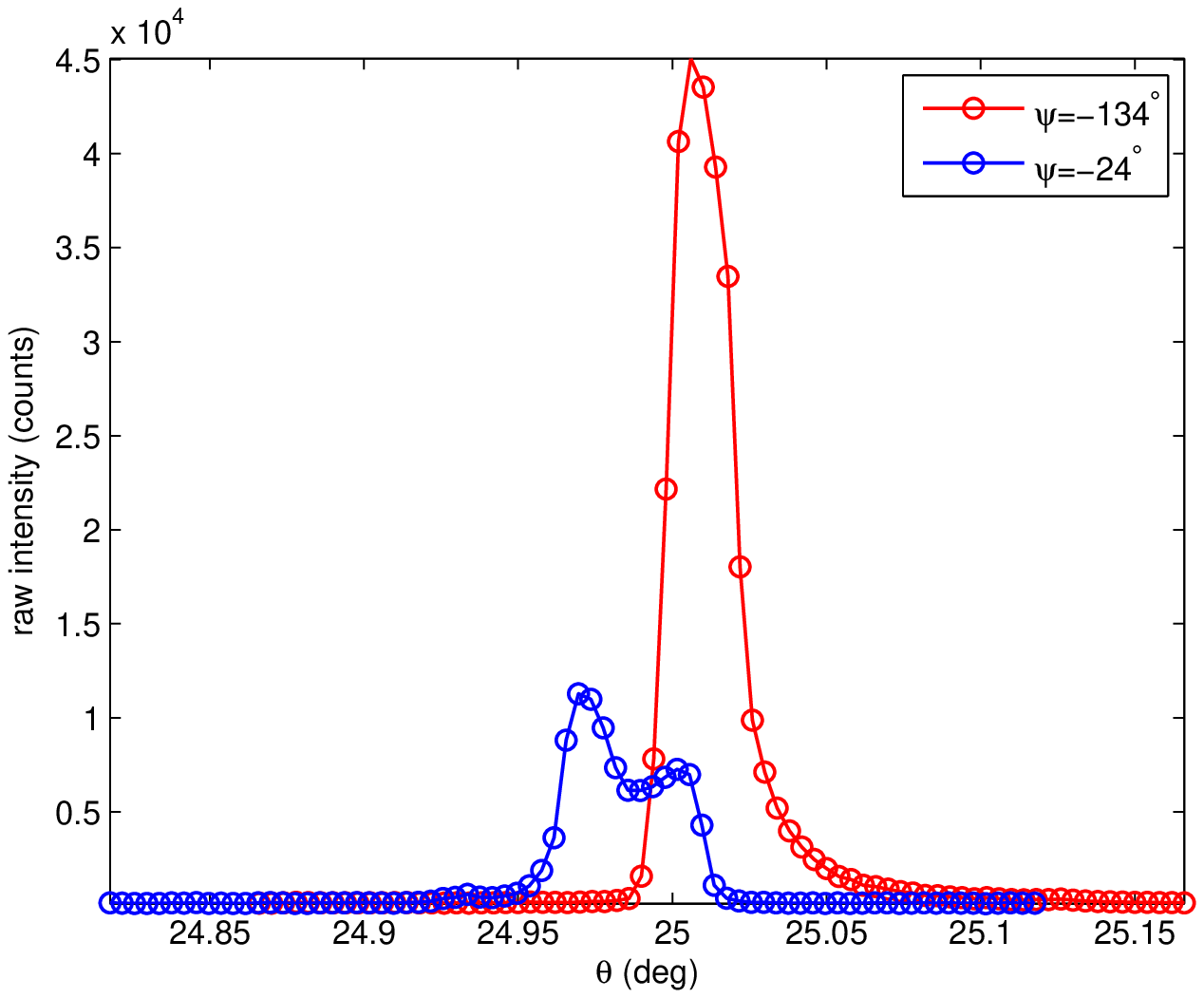}
    \includegraphics*[width=0.49\columnwidth]{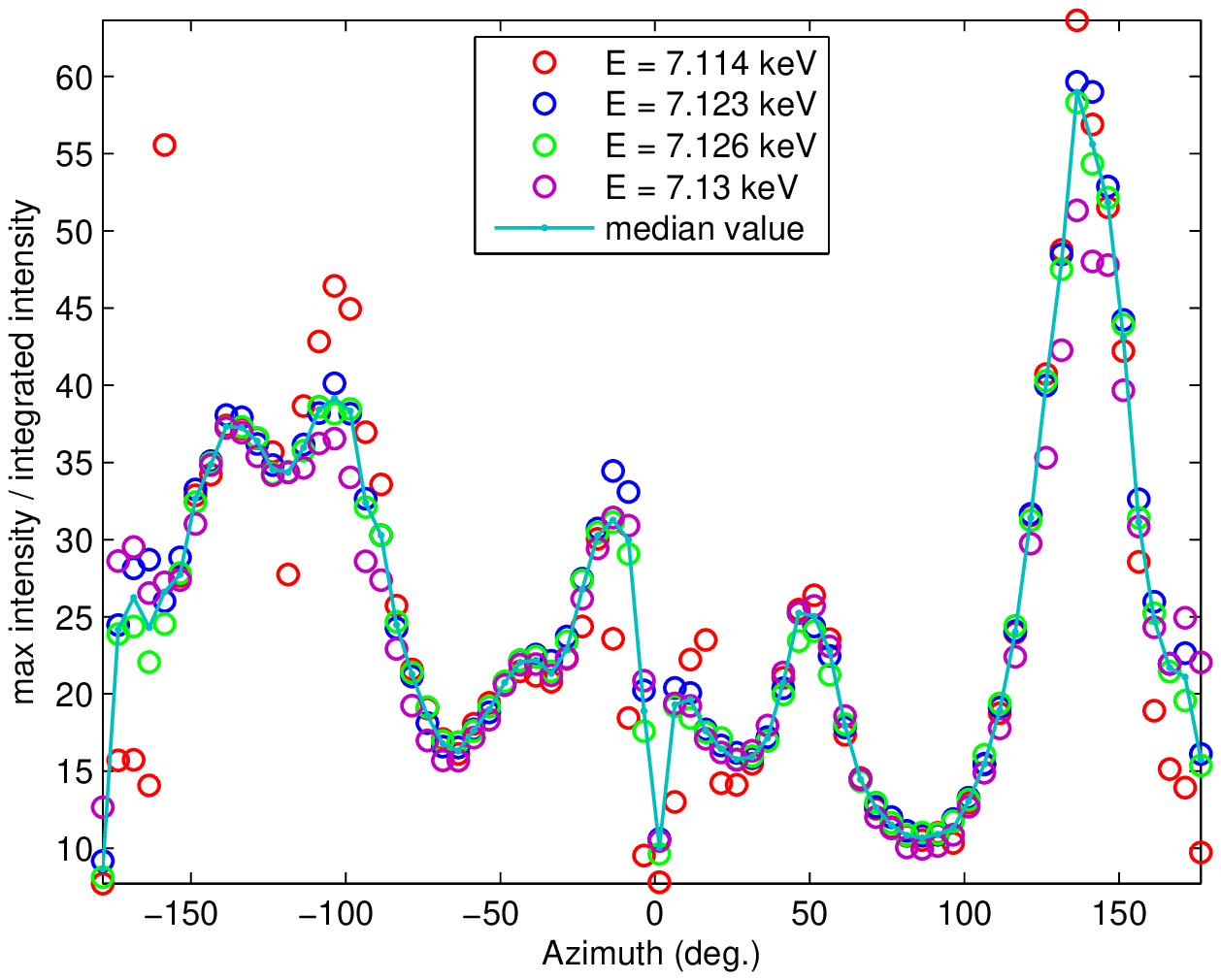}
    \caption{Left: rocking curves of the reflection $500$ at two different azimuths, illustrating the variability of the shape (hence the crystal quality). Right: maximum intensity to integrated intensity ratio  of the $500$ reflection rocking curves versus azimuth. The intensities recorded at 4 different energies show that this ratio  is essentially energy independent.}
    \label{fig.hiratio}
\end{figure}

Another problem is the inhomogeneity of the sample, which leads to strong variations of the  rocking curve profiles obtained in  azimuthal scans: the ratio of the  maximum height of the rocking curve to its integrated intensity is not constant. Since the energy-azimuth maps were recorded "on the fly", i. e. without integrating each rocking curve, we had to correct for the azimuthal dependence of the maximum/integrated intensity ratio of rocking curves,  that were recorded at a few sample energies (Fig. \ref{fig.hiratio}). As expected, these  ratios seem to be energy independent  within the  small energy range of the spectra.

\begin{figure}[ht]
    \includegraphics*[width=0.49\columnwidth]{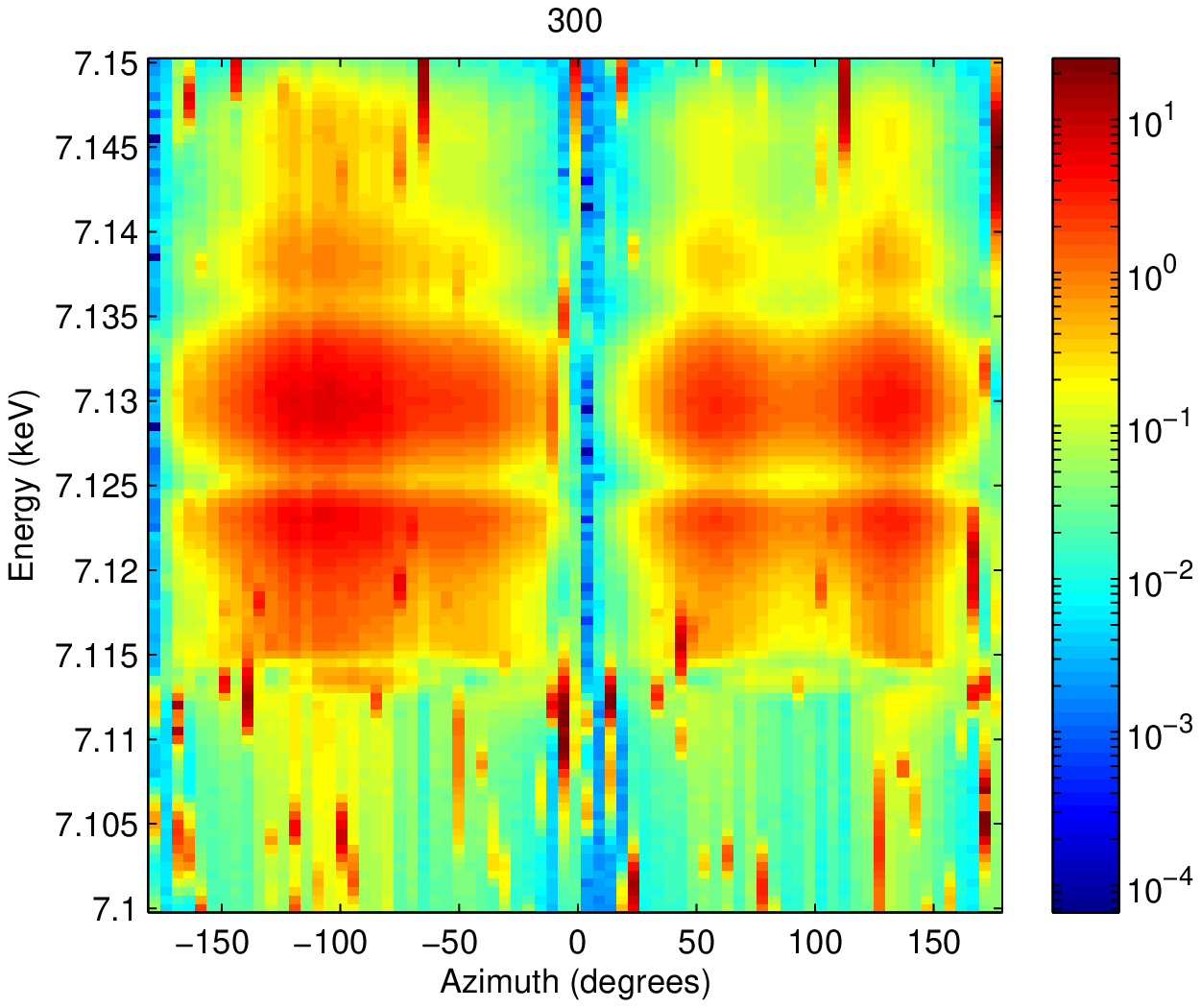}
    \includegraphics*[width=0.49\columnwidth]{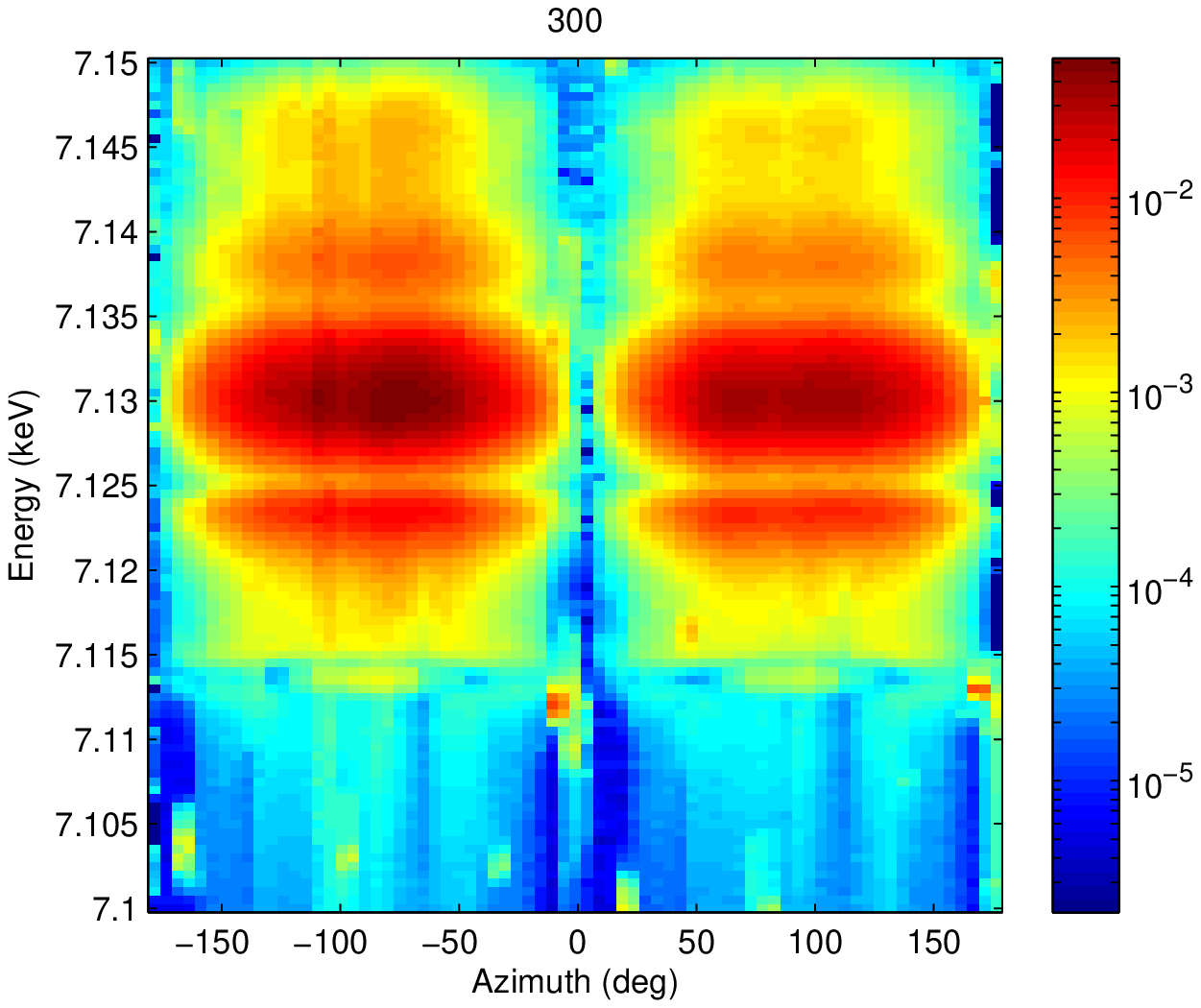}
    \includegraphics*[width=0.49\columnwidth]{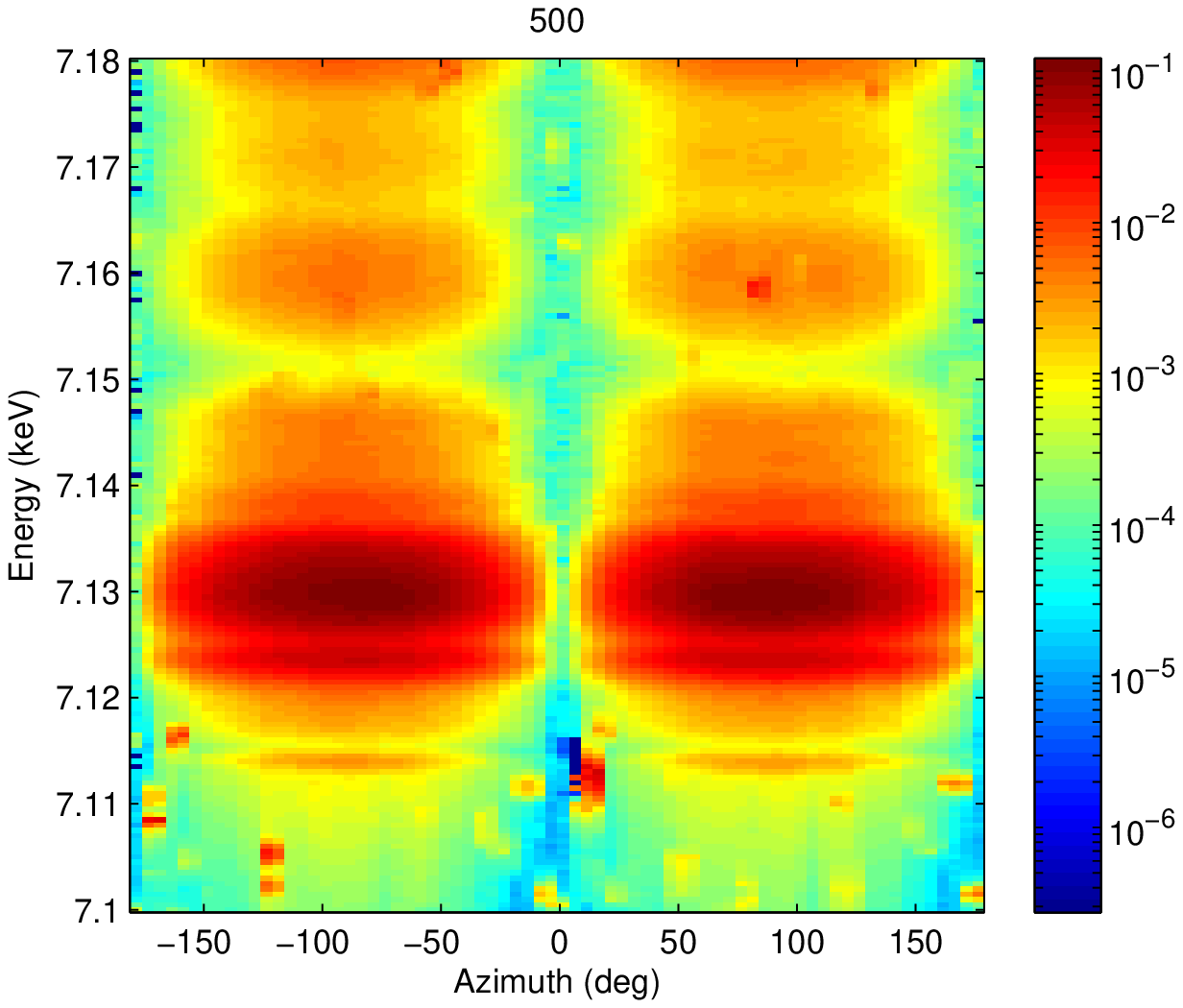}
    \includegraphics*[width=0.49\columnwidth]{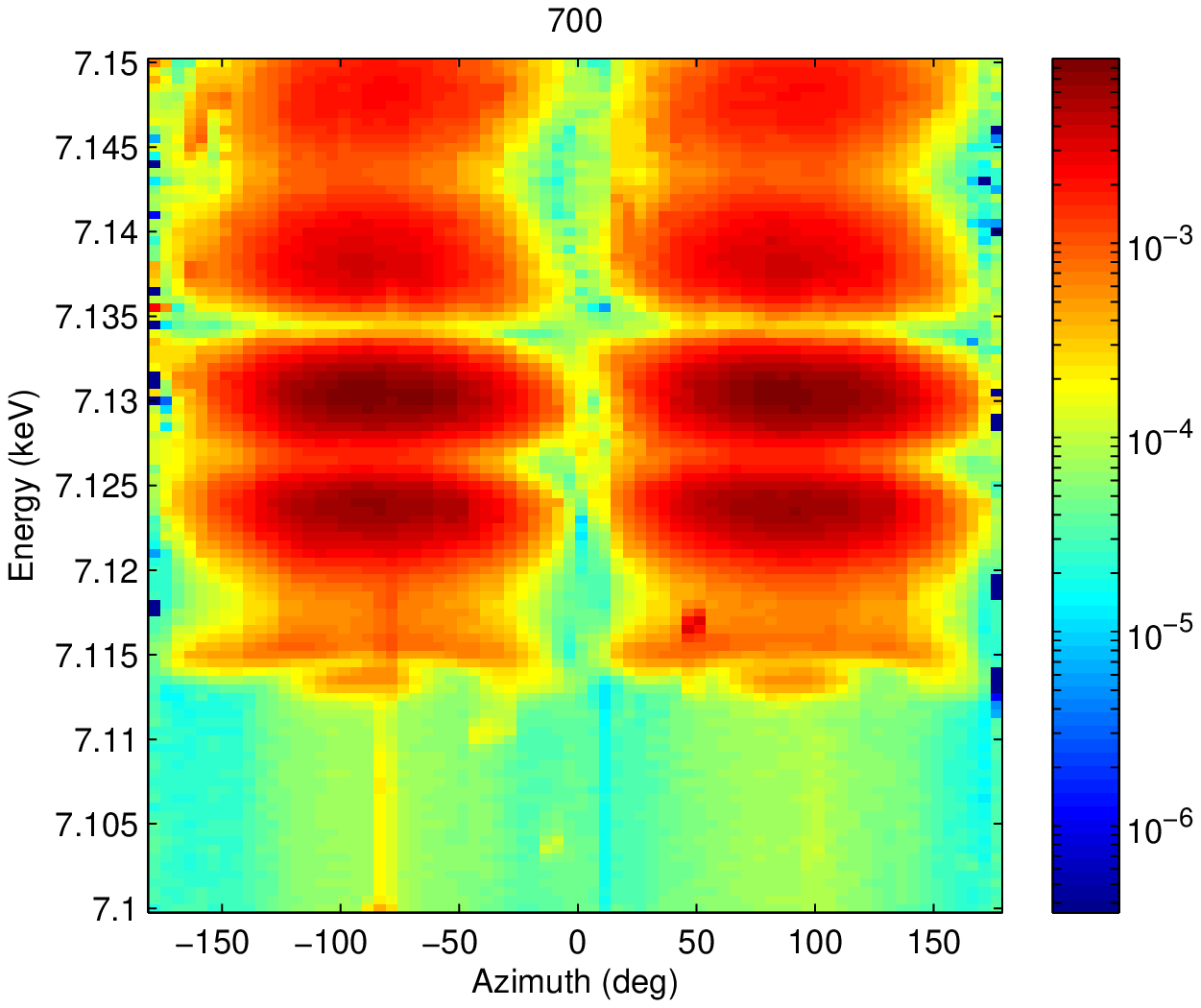}
    \caption{Energy-azimuth maps of the $300$ (top), $500$ (bottom left) and $700$ (bottom right) forbidden reflections. For the $300$, both raw data (left) and corrected data (right) are shown. Note the distortion of the azimuthal dependence by the poor crystal quality, the distortion of the energy dependence by the self-absorption, and the dense multiple scattering. For the $500$ and $700$, only corrected data are shown.}
    \label{fig.maps}
\end{figure}

The last effect to correct for is multiple scattering. In fact, most of the energy spectra are severely contaminated by multiple scattering (Fig \ref{fig.maps}, top left). To correct for this, we eliminated all experimental points whose intensities  relative to those of their azimuthal neighbors  exceeded a threshold value which was chosen as the standard deviation of the next-neighbour intensity variation. The  eliminated points were then replaced by interpolations from the remaining points. This method takes advantage of the small azimuthal step size (5$^\circ$) of the experimental data and  most of the multiple scattering could be removed, the remaining contributions being too small to have significant impact on subsequent data analysis (Fig \ref{fig.maps}, top right).

The resulting two-dimensional azimuth-energy maps of intensities of the forbidden reflections 300, 500 and 700 are shown in Figure \ref{fig.maps}.

\begin{figure}[ht]
    \includegraphics*[width=0.32\columnwidth]{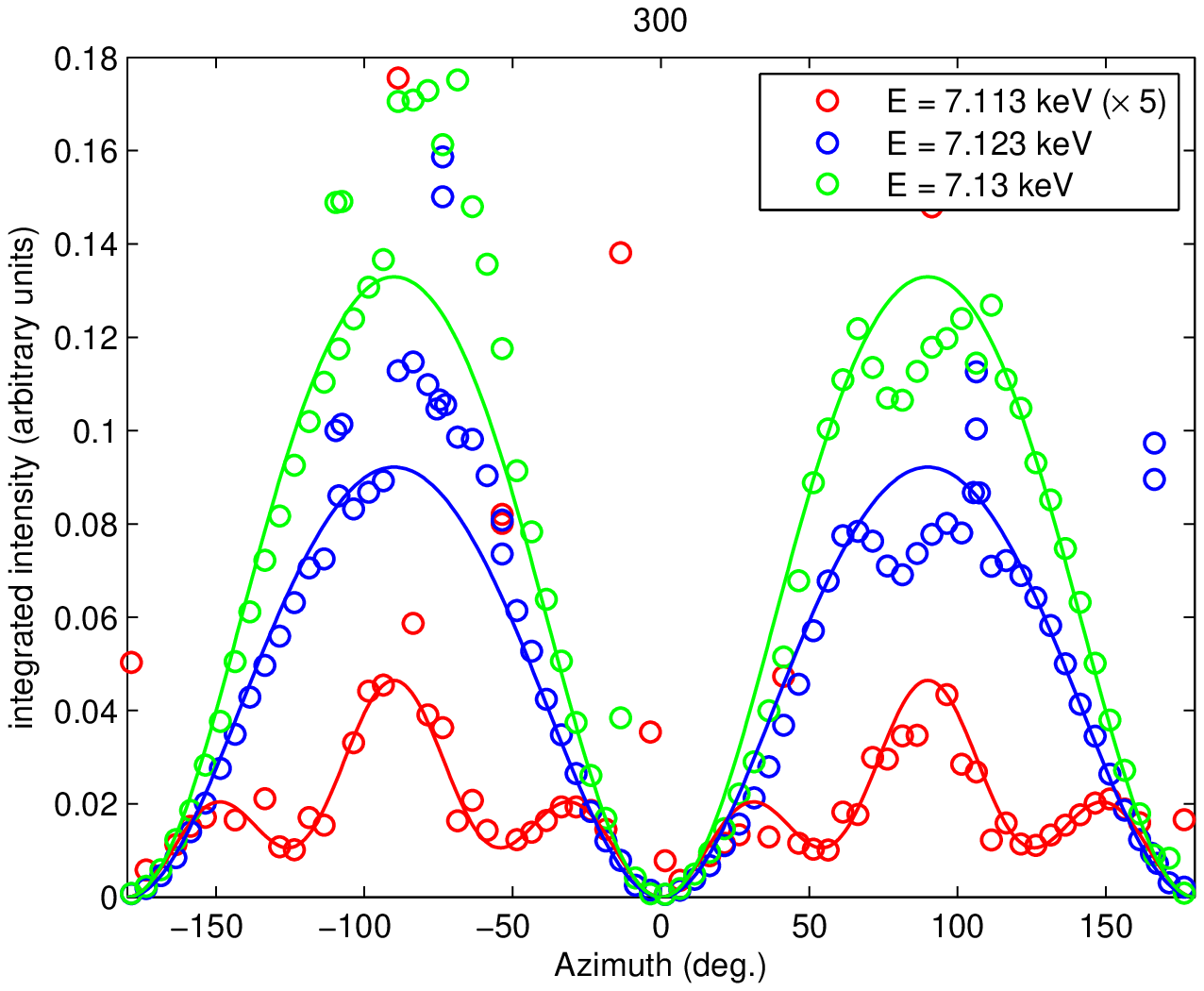}
    \includegraphics*[width=0.32\columnwidth]{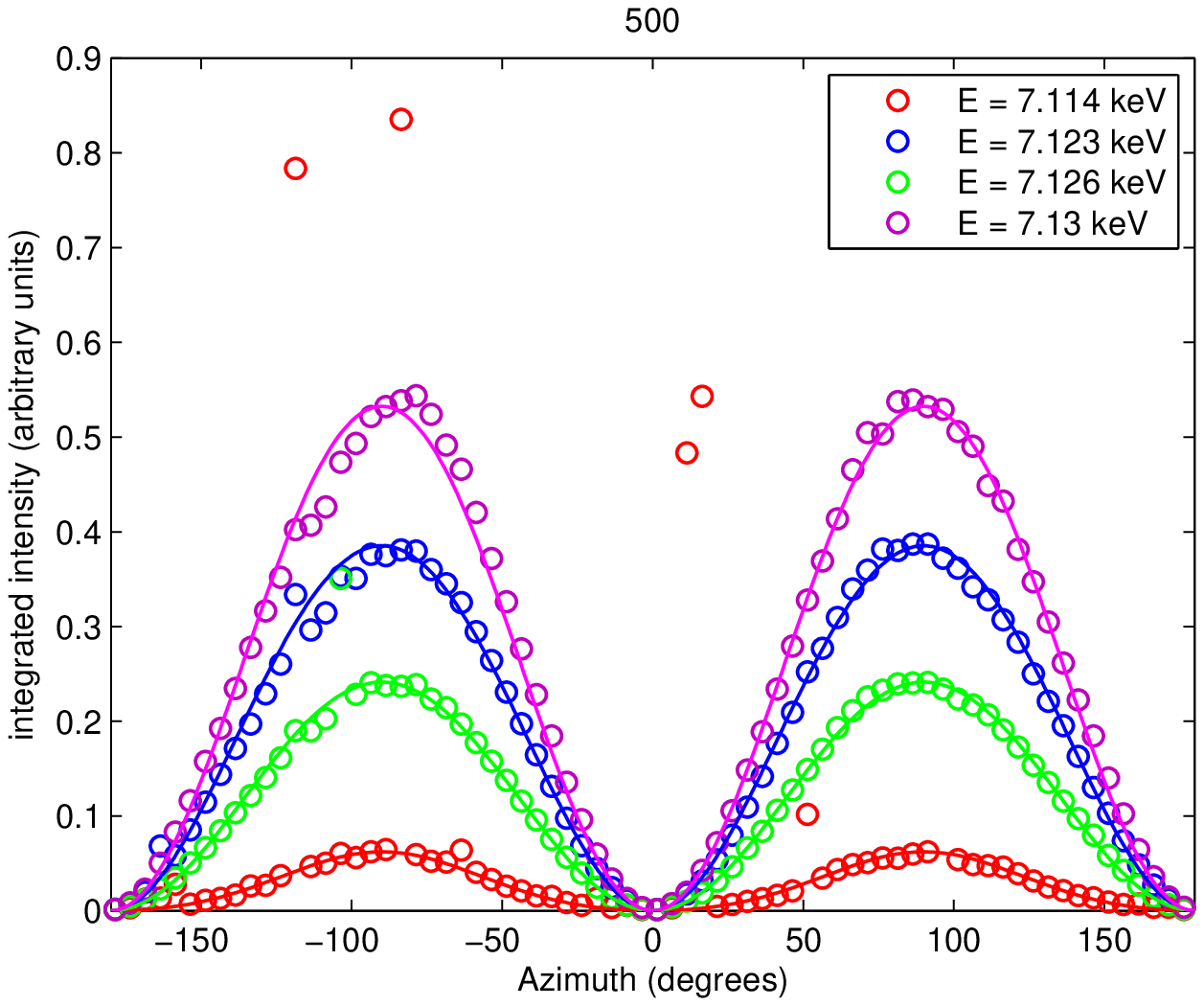}
    \includegraphics*[width=0.32\columnwidth]{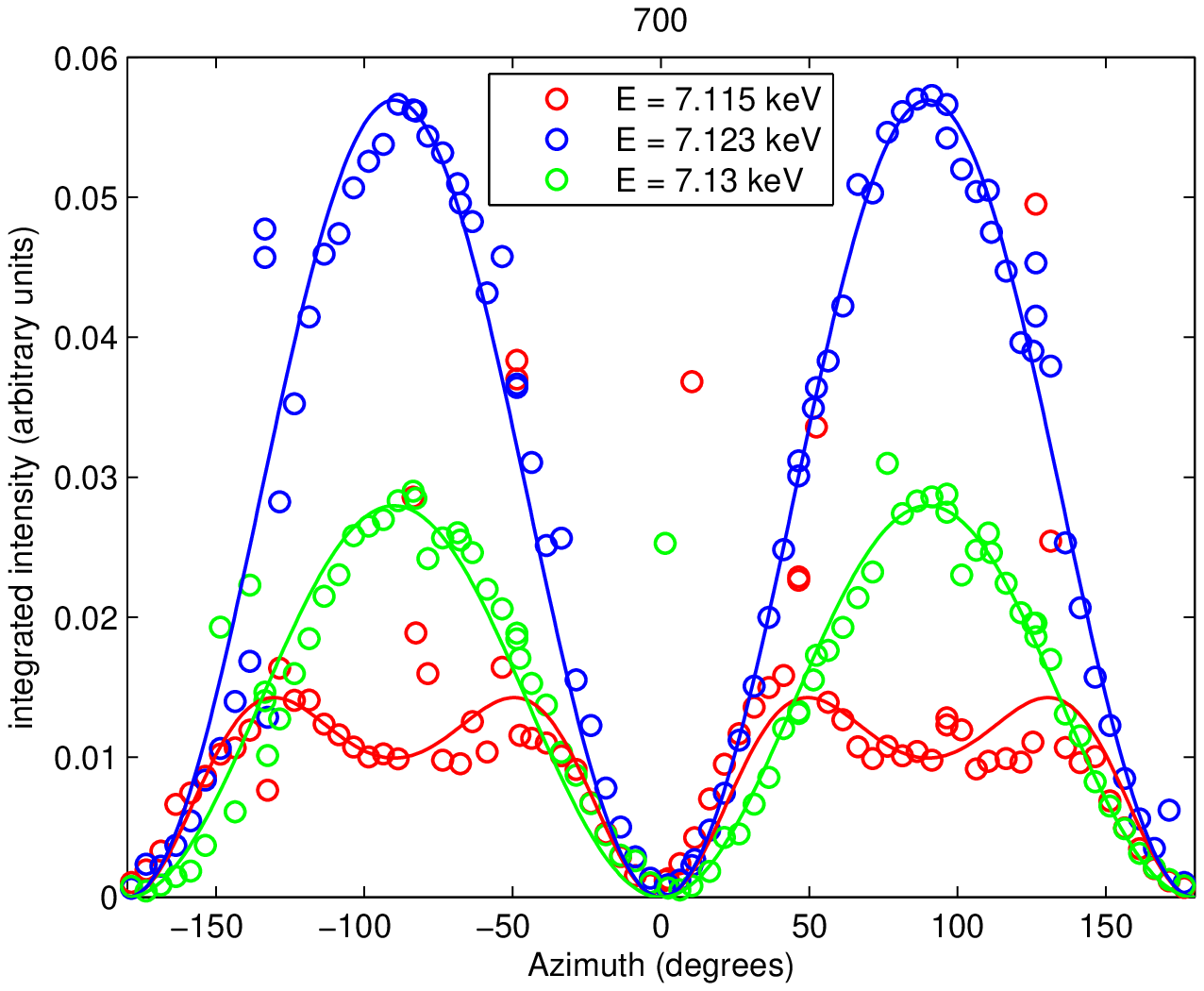}
    \includegraphics*[width=0.32\columnwidth]{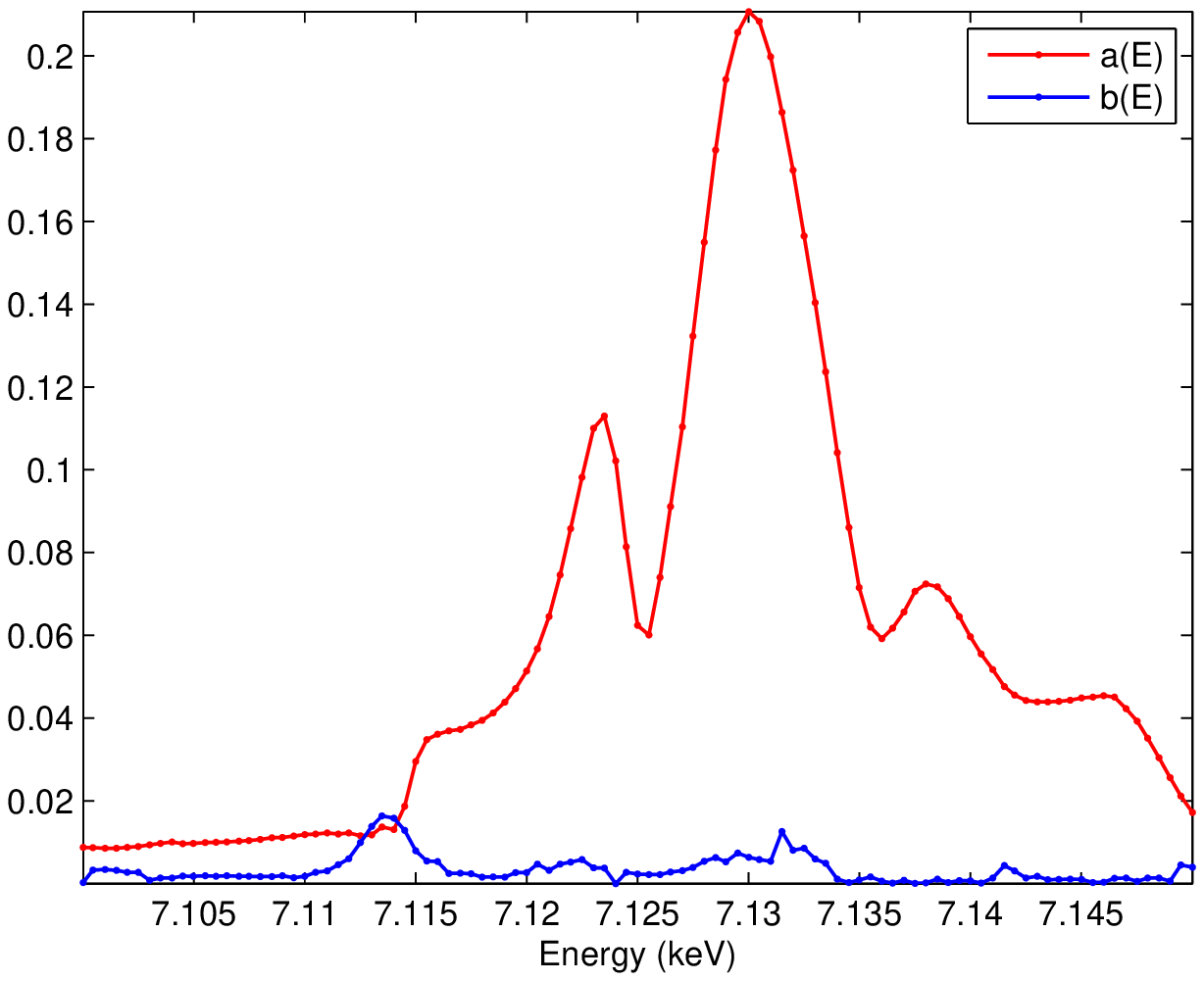}
    \includegraphics*[width=0.32\columnwidth]{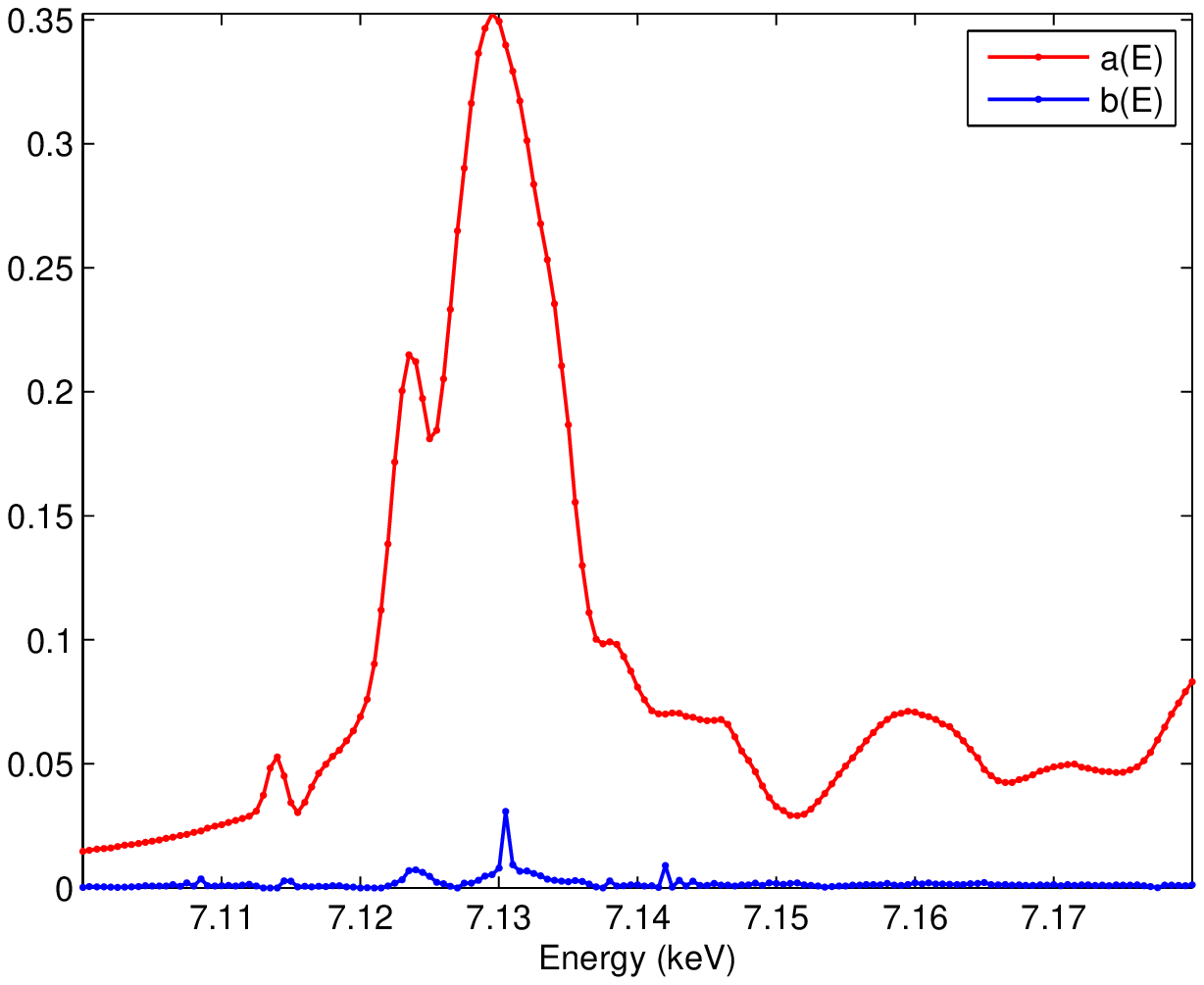}
    \includegraphics*[width=0.32\columnwidth]{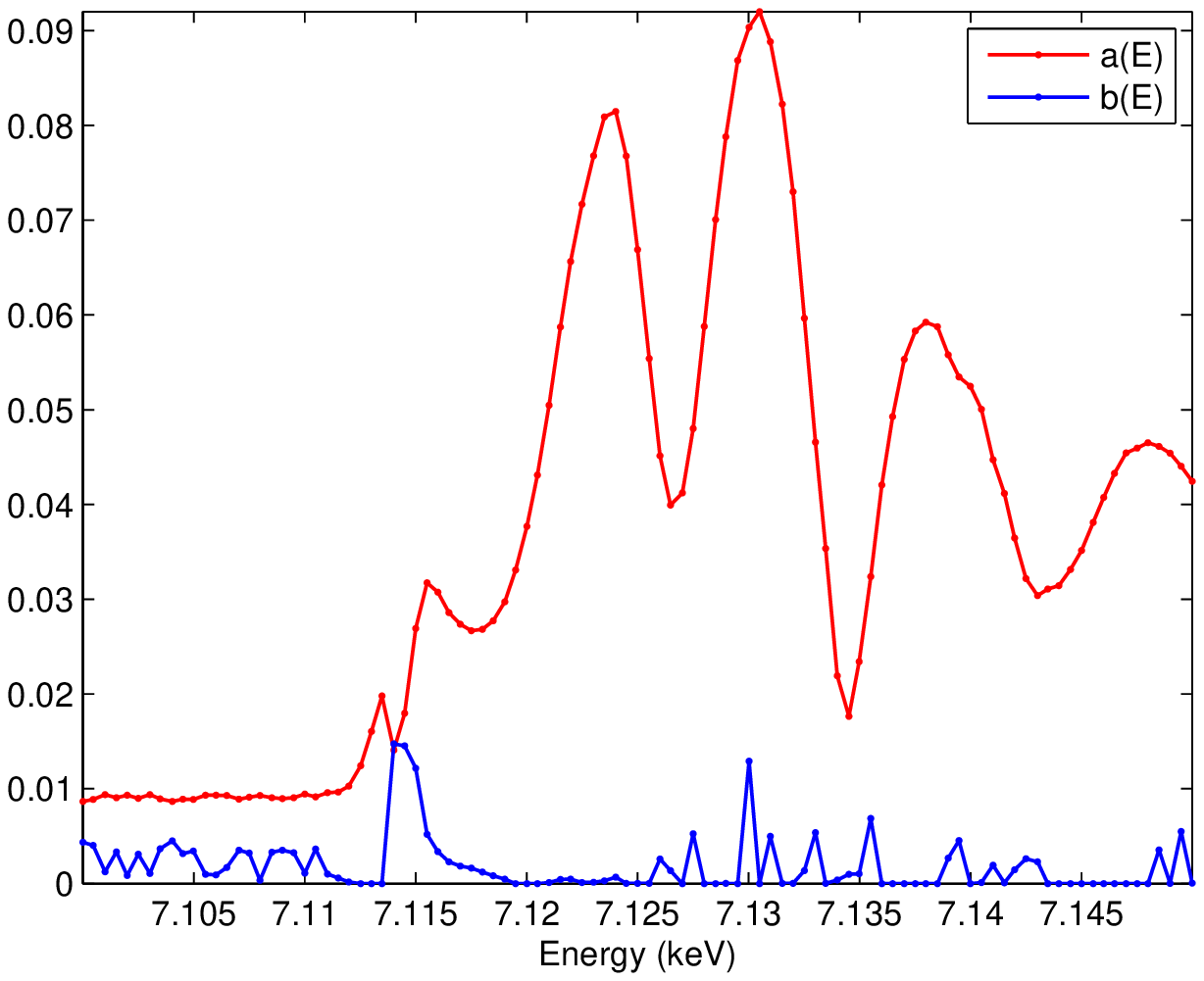}
    \caption{Top: selected azimuthal dependences and their fits against equation (\ref{eq.rxsfit}). A hint of asymmetry, visible in the pre-edge of reflections $300$ and $700$, can be attributed to non-resonant magnetic scattering. Bottom: fit parameters $a(E)$ and $b(E)$ of the energy-azimuth maps of Figure \ref{fig.maps} with the model given in equation (\ref{eq.rxsfit}). Left: 300, middle: 500, right: 700.}
    \label{fig.rxsfit}
\end{figure}

In a further step,  according to the theoretical predictions (Eq. (\ref{eq.ddsp}), (\ref{eq.dqsp}) and (\ref{eq.qqsp})), the corrected data $I(E,\psi)$ were fitted with the model function:
\begin{equation}
    I(E,\psi)=\left|a(E)\sin\psi+b(E)e^{i\Phi(E)}\sin(3\psi)\right|^2
    \label{eq.rxsfit}
\end{equation}
where $a(E)$, $b(E)$ and $\Phi(E)$ are real positive parameters (Fig. \ref{fig.rxsfit}).
The differences between the 'corrected' maps and the fitted maps are  essentially due to the residual multiple scattering.
As expected, the fit curves show a strong 2-fold component, mainly due to the dipole-dipole scattering which largely hides
 the 2-fold dipole-quadrupole scattering as well as the 2-fold part of the quadrupole-quadrupole scattering.
The 4-fold component of the  quadrupole-quadrupole scattering appears only in the pre-edge region as expected.
The small features at higher energies may be explained by the isotropic absorption correction: the 2-fold aspects of the scattering ($\left|\sin\psi\right|^2$) and of the absorption ($\cos2\psi$) yield a weak 4-fold contribution, like the interference between the $£\sin\psi$ and $\sin3\psi$ terms in the scattering amplitude. This contribution appears on the main peaks of the reflections 300 and 500, where both intensity and absorption are large. Elsewhere it is within the noise of the measurement. Its weakness proves that our approximation of  isotropic absorption suffices for the correction of the RXS data. The phase parameter $\Phi(E)$ remains close to $0[\pi]$, except for  reflection 700 for which the 4-fold feature seems to  be shifted in phase by $\pi/2[\pi]$ against  the 2-fold one. However, the data for  the fit are too noisy to state this with confidence.

\subsection{Thermal-Motion-Induced scattering and magnetic scattering}

\begin{figure}[ht]
    \includegraphics*[width=0.49\columnwidth]{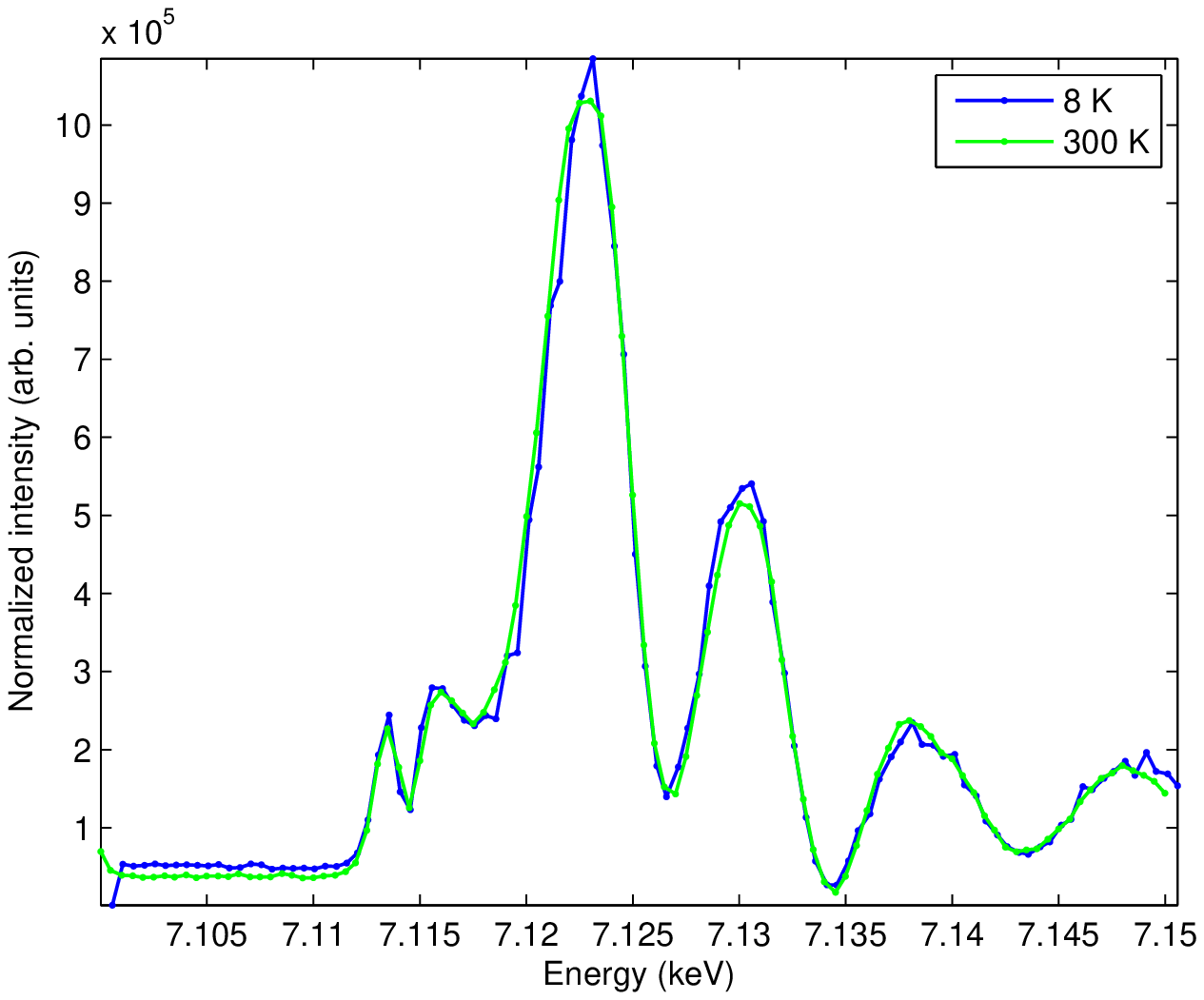}
    \includegraphics*[width=0.49\columnwidth]{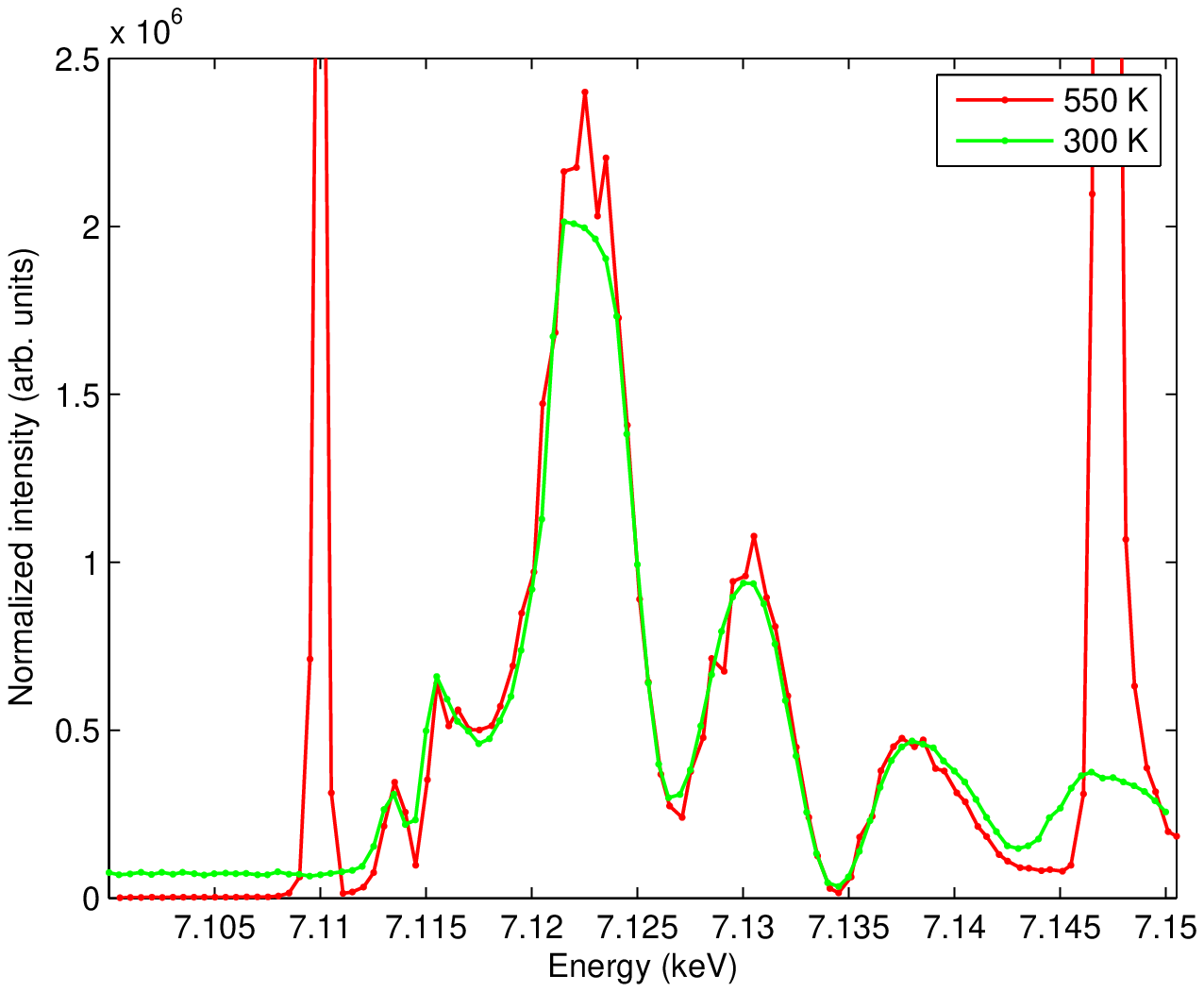}
    \caption{Energy spectra of the forbidden reflection $700$ at low temperature (8 K), room temperature (300 K) and elevated temperature (550 K). The normalized intensities without absorption corrections are presented. Note that the data at room temperature were taken at XMaS, whereas  the data 8 and 550 K were measured at DLS I16. Left: Azimuth $\psi=90^\circ$. Right: Azimuth $\psi=-72^\circ=-90^\circ+18^\circ$ at 550 K and $\psi=-108^\circ=-90^\circ-18^\circ$ at 300 K.}
    \label{fig.temp}
\end{figure}

So far,  our analysis is based on the hypothesis that the forbidden
reflections are exclusively due to time-even scattering of the atoms in
their nominal positions, i.e. we neglect contributions of
Thermal Motion Induced Scattering (TMI).
TMI is the derivative of the scattering tensor of the atom at rest with respect to its position.
Due to the predominating  dipole-dipole scattering, its main contribution
is therefore a third rank tensor with the same symmetry
constraints as those of the symmetric dipole-quadrupole scattering tensor, and, being
related to thermal displacements of atoms, TMI almost vanishes at low temperatures.
Therefore, we measured the reflection $700$ also at low temperature.
The energy spectra at room temperature (300 K) and at low temperature (8 K) are almost identical (Fig. \ref{fig.temp}, left).
This finding proves that TMI  must be small compared to the tensor of scattering at rest.

Also so far, we have considered the crystal as non magnetic although it is
weakly ferromagnetic with a paramagnetic transition at $T_c=508$ K.
Despite of the small ferromagneticity, most of
its ordered moments follow an antiferromagnetic order with
propagation vector $[000]$: the spins form two ferromagnetic
sub-layers in the ($a$,$b$)-plane, and the coupling between the
two sub-layers, along the $c$-axis, is antiferromagnetic \cite{maltsev}.
It undergoes a spin reorientation transition at
$T_{SR}=415$ K, when the antiferromagnetic and ferromagnetic axes
swap their directions: the antiferromagnetic axis goes from [001]
below $T_{SR}$ to [100] above $T_{SR}$ \cite{wolfe}.
As a consequence of the antiferromagnetic component, the magnetic structure factor
does not vanish for the forbidden reflections $h00, h=2n+1$, i.e.
both  resonant and non-resonant magnetic scattering can contribute
to their intensities.

The magnetic moments are carried by the Fe$^{3+}$ cations, which, in iron oxides,  are normally in high spin state ($S=5/2$ and $L=0$) \cite{iron_oxides}.
Since resonant electric transitions are only sensitive to  orbital moments
(and to the spin moments  via  spin-orbit coupling),  a resonant magnetic scattering contribution to the forbidden reflections  can be expected  very weak, but
a contribution of  non-resonant magnetic scattering cannot be ruled out.
Indeed, the experimental spectra derived at room temperature and low temperature do not vanish
below the absorption edge, which suggests measurable effects of  non-resonant magnetic
scattering (Figs. \ref{fig.rxsfit} and \ref{fig.temp}).
This assumption of non-resonant magnetic
scattering  is confirmed by the  measurements performed  at 550 K, i.e. well above the paramagnetic
transition, which failed to yield any measurable signals below the
edge (Fig. \ref{fig.temp}, right). Close to and above the
edge, the strong absorption reduces the scattering volume in such way
that non-resonant contributions become small compared to the
resonance enhanced contributions.
Since our aim is to observe and demonstrate  the
interferences of the RXS from the two iron sites, we will not further
focus on the non-resonant region and  ignore  magnetic
scattering in the ensuing data analysis.

\section{Fitting the energy and azimuthal dependencies}
\label{fit}

Our symmetry analysis has shown that the structure factors of the $h00, h=2n+1$ reflections
can be conveniently described by 12 tensor components (6 for each site),
but azimuthal scans allow only for extracting 2 independent spectra per reflection.
Therefore, the experimental results can only be interpreted by numerical simulations.
To this purpose, the FDMNES code \cite{fdmnes} was used for calculating  the absorption cross-sections and the structure factors $|F(\ten H)|^2$ of the forbidden reflections. This information served to yield data that can
be compared with the experimental intensities, after correction.
Using  the multiple scattering option of the code, with the Hedin-Lundquist potential, it was attempted to optimize the model  parameters with respect to best agreements with  the experimental values of $I(\ten H)$ and $\mu(E)$.

Let us refer to the strong reflections above
7.120 $keV$ as main peaks and to the weak peaks below 7.120 $keV$ as pre-edge peaks.
The numerical simulation of the various contributions to the tensor atomic factors show that while the main peaks are mainly due to the dipole-dipole
contributions from  both of the iron cations, the pre-edge peaks can be
described by higher order terms producing complex
azimuthal and energy structures.
Despite of the dipole-dipole origin of the main peaks, their intensity modelling is not simple, just because of
the interference of the radiation scattered at the two Fe positions.
It follows from Table \ref{phases}, that $4(c)$ and $8(d)$ structure factors
interfere in very different ways for the $300$, $500$ and $700$ reflections.
Indeed, the experimentally observed $300$ and $500$ reflections are strong,  contrary to the weak $700$,
and the $700$ energy spectrum looks rather different from those  of  $300$ and $500$.

\begin{figure}[ht]
    \includegraphics*[width=0.49\columnwidth]{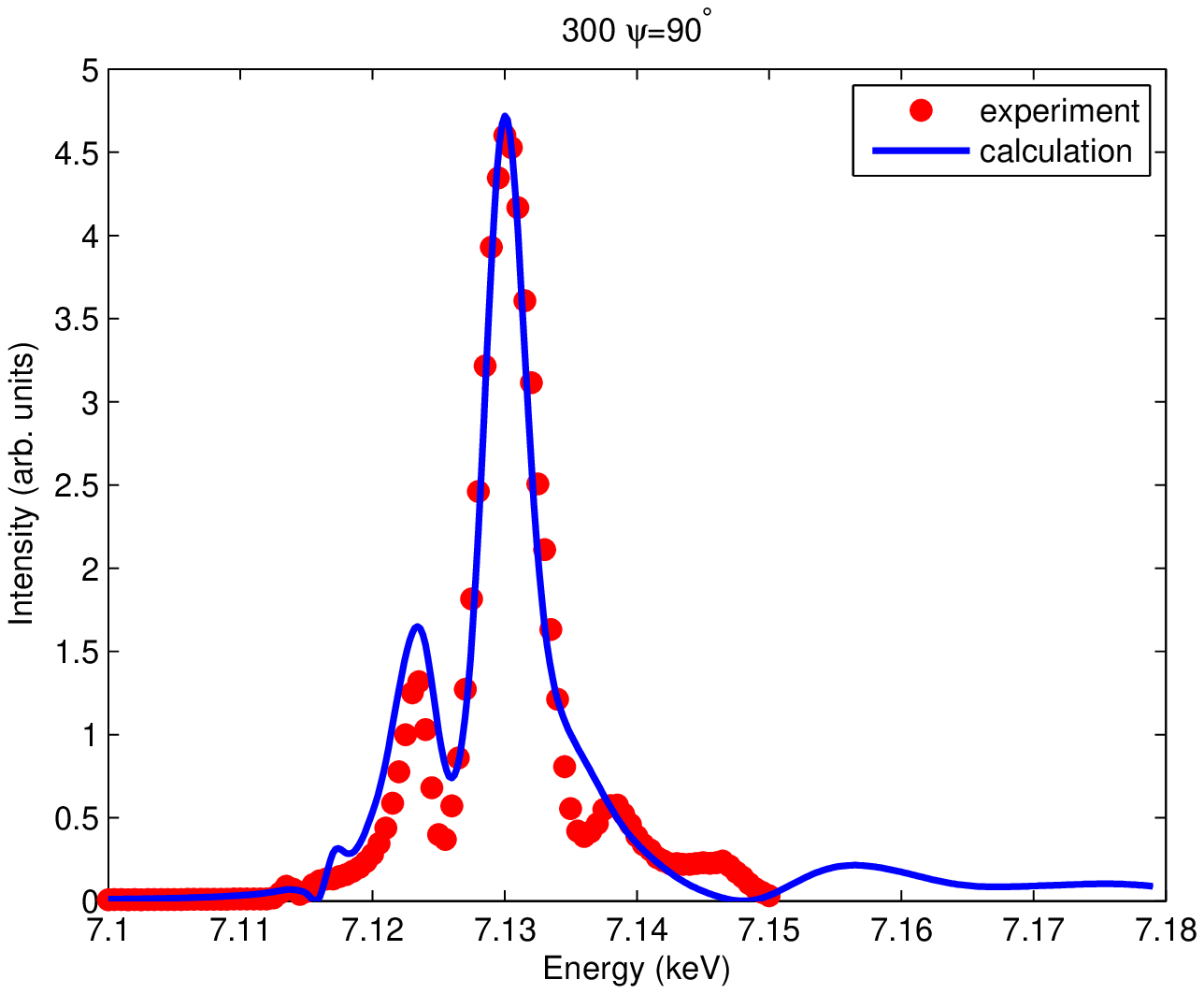}
    \includegraphics*[width=0.49\columnwidth]{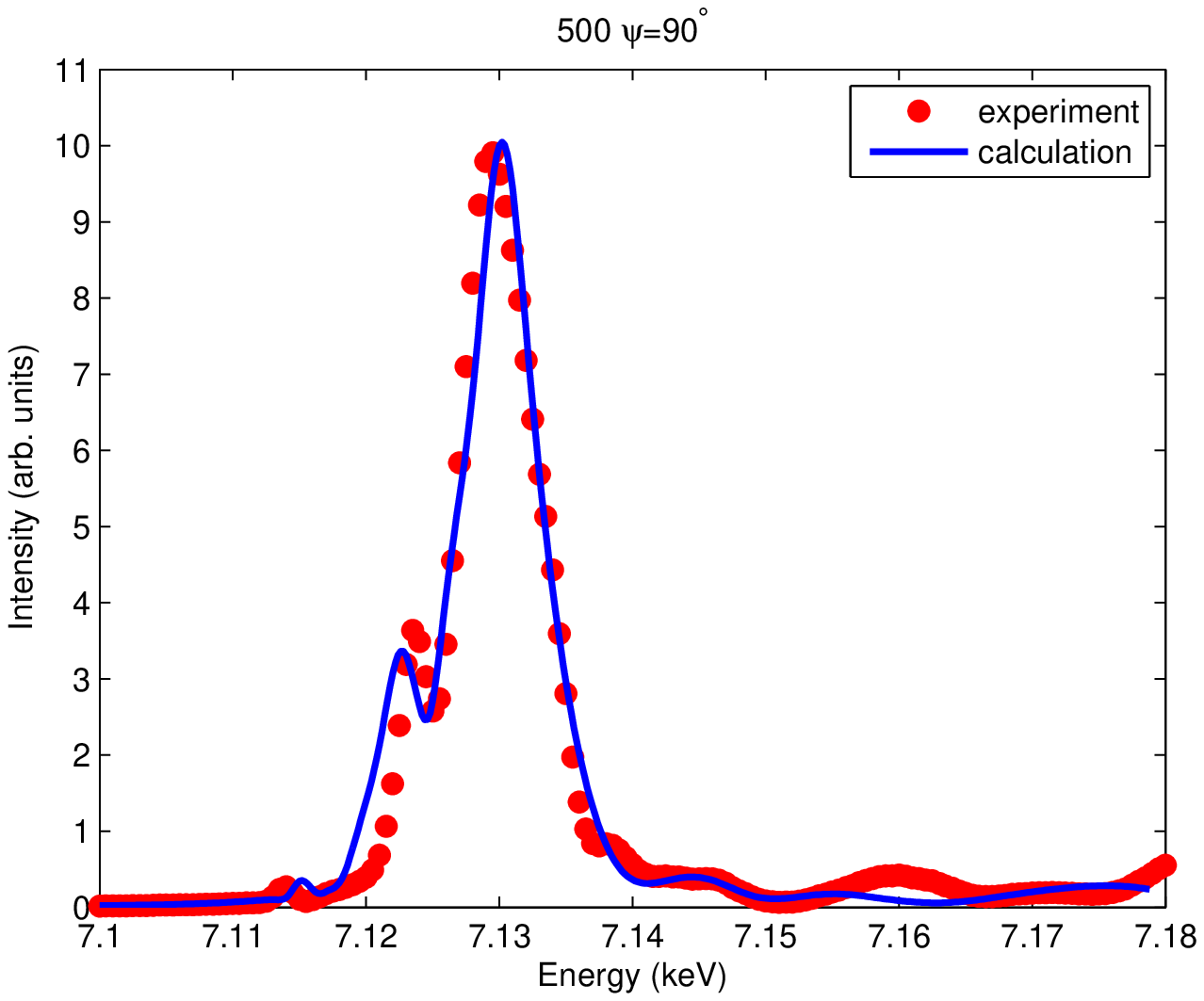}
    \includegraphics*[width=0.49\columnwidth]{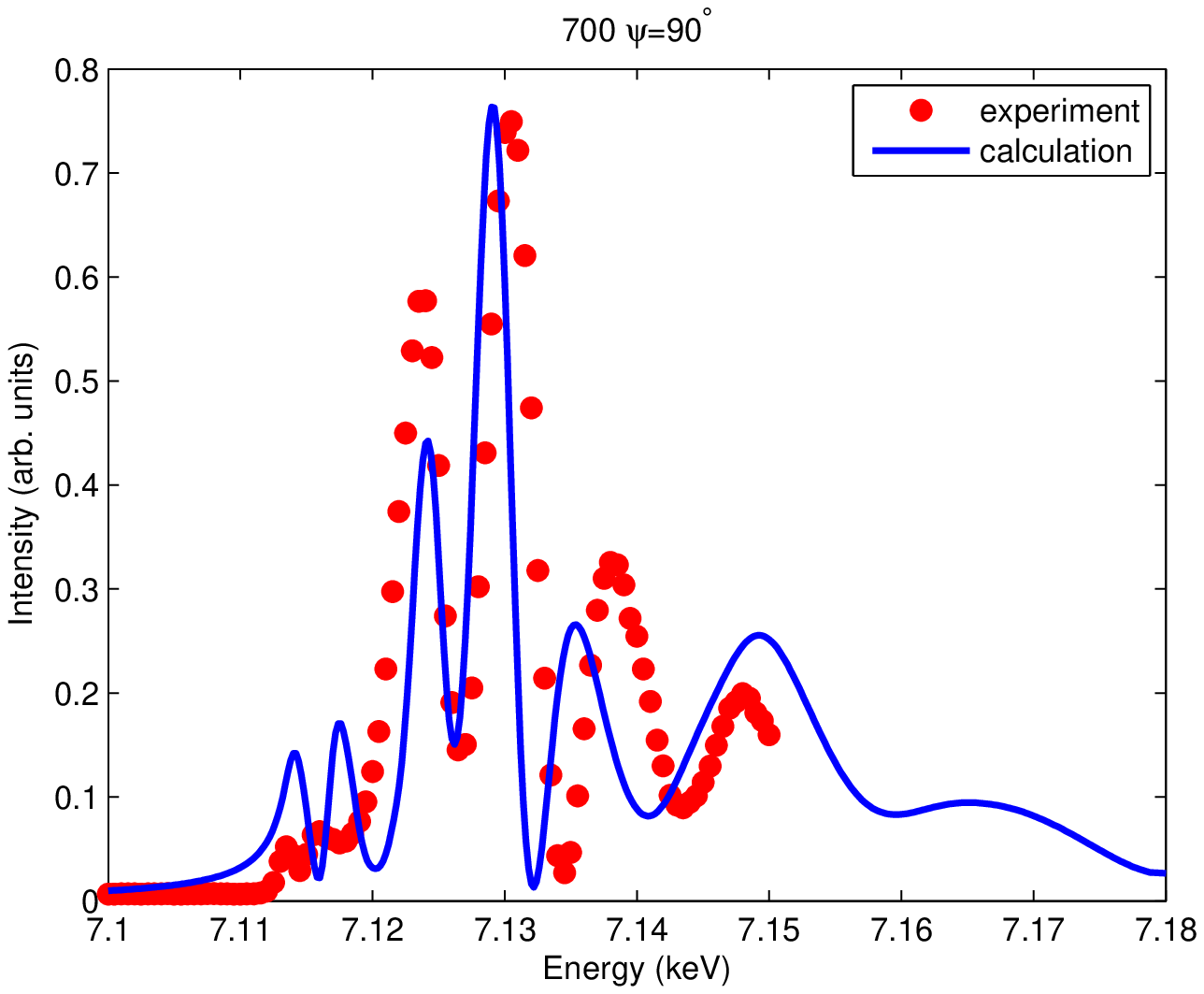}
    \caption{Energy spectra of the reflections $300$, $500$ and $700$ at the azimuth $\psi=90^\circ$: FDMNES calculations versus experimental data.}
    \label{fig.spectra}
\end{figure}

In the course of simulations,  the main peaks of the 3 measured reflection intensities were fitted first.
Neglecting higher order terms in their energy range, the model intensities depend on two parameters only: $D_{xz}^{8(d)}$ and $D_{xz}^{4(c)}$.
Table \ref{phases} shows that
the dipole-dipole component of the $300$ intensity is mainly provided by the $8(d)$ sites,
while the contribution of the $4(c)$ sites is very small.
This circumstance allows for a convenient extraction of $D_{xz}^{8(d)}$,  whereupon $D_{xz}^{4(c)}$ could be extracted from the $500$ spectrum, once $D_{xz}^{8(d)}$ was known.
At last, the $700$, in which the radiation interference between the iron sites is mainly destructive, was used for fine tuning, because the destructive interference makes modelling  extremely sensitive to the ratio of both parameters.
A  satisfactory description of the $700$ spectrum could only be achieved upon introducing an energy  shift of 0.7 $eV$ between both contributions.
This chemical shift can be rationalized  by the different environments of the iron cations  occupying two inequivalent sites.
Then, it was checked that the same set of parameters allows for reasonably good 'fits' of all 3 reflections (Fig. \ref{fig.spectra}) as well as  of the absorption (Fig. \ref{fig.fluo}).

The key step  of the simulation procedure is the convolution that describes the spectral broadening $\Gamma(E)$, due to the core- and valence electron lifetimes in the excited state.
This function being  unknown, FDMNES supposes $\Gamma(E)$ to  grow smoothly, e.g. like
\begin{equation}
\Gamma=\Gamma_{hole}+\Gamma_m \left[\frac 1 2
+\frac{1}{\pi}\arctan\left(\frac{\pi}{3}\frac{\Gamma_m}{E_{large}}\left(e-\frac{1}{e^2}\right)\right)\right]
\end{equation}
where $e=\frac{E-E_{Fermi}}{E_{cent}}$.
The 5 parameters $\Gamma_{hole}$, $\Gamma_m$, $E_{cent}$, $E_{Fermi}$, and $E_{large}$,
that provided the least differences from the experimental data, were obtained with the optimization procedure of the FDMNES code, yielding: $E_{Fermi}=-6.5 eV$, $\Gamma_m=24 eV$,
$E_{cent}=14 eV$, $E_{large}=10 eV$, $\Gamma_{hole}=2 eV$.

\begin{figure}[ht]
    \includegraphics*[width=0.49\columnwidth]{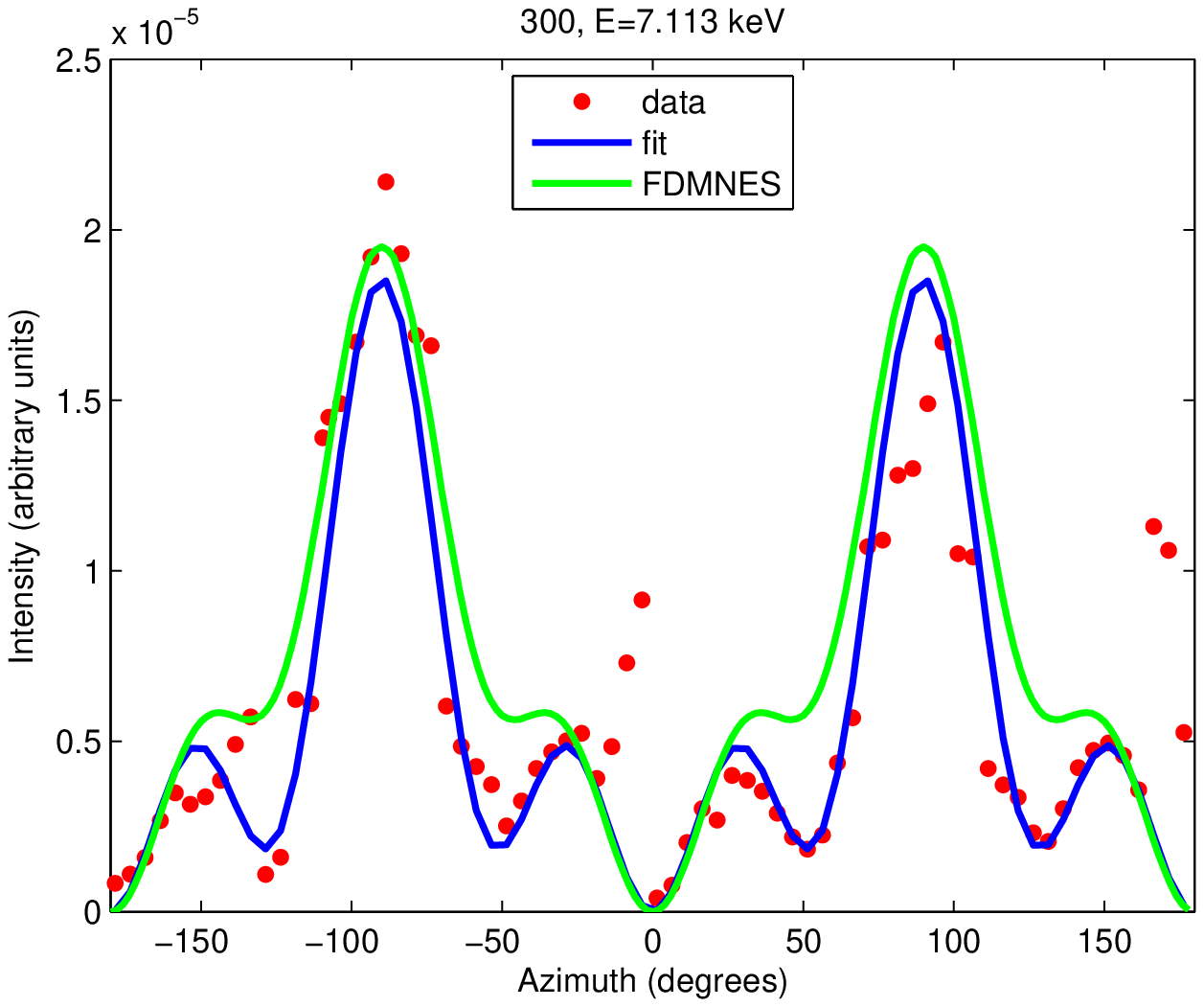}
    \includegraphics*[width=0.49\columnwidth]{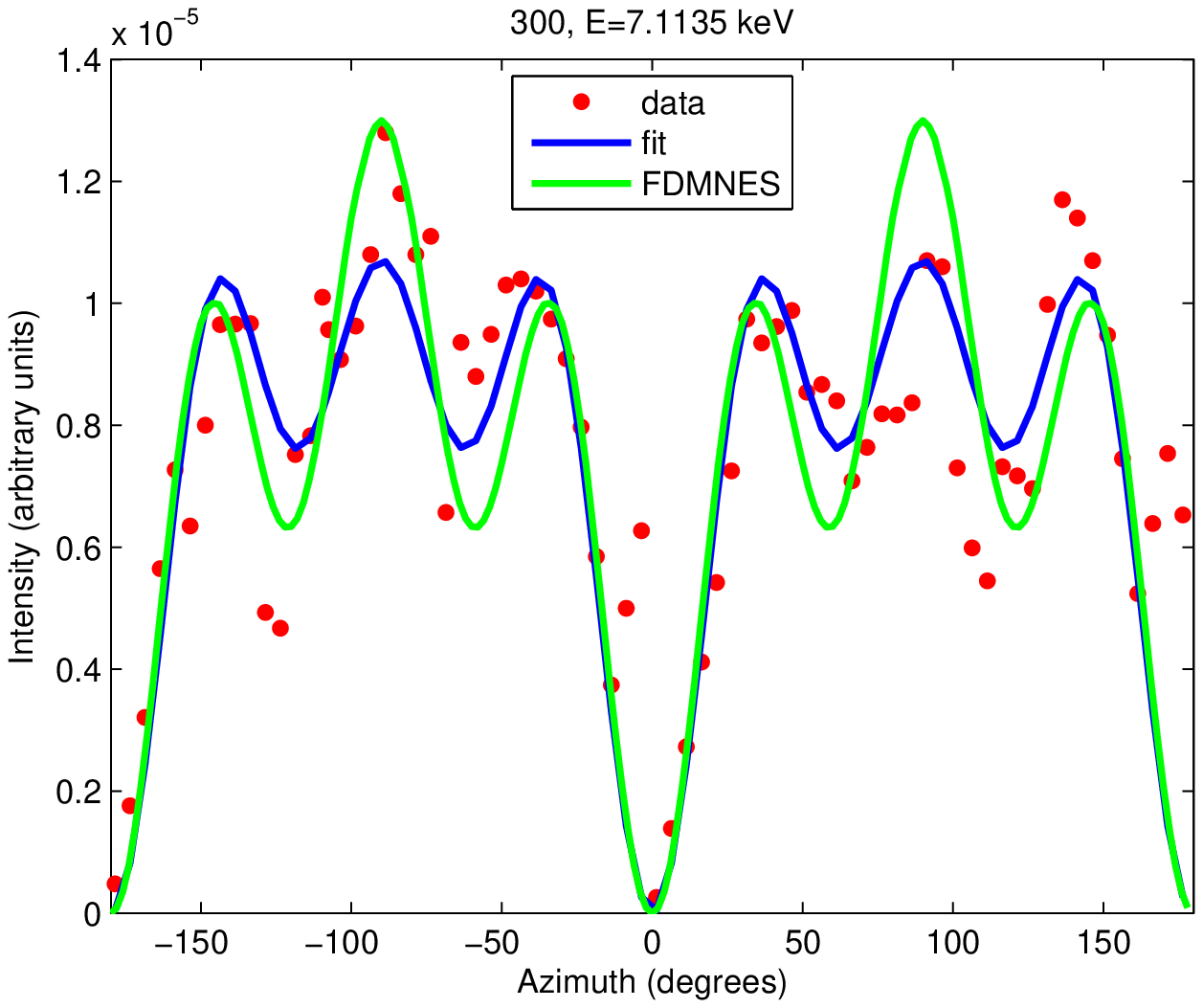}
    \includegraphics*[width=0.49\columnwidth]{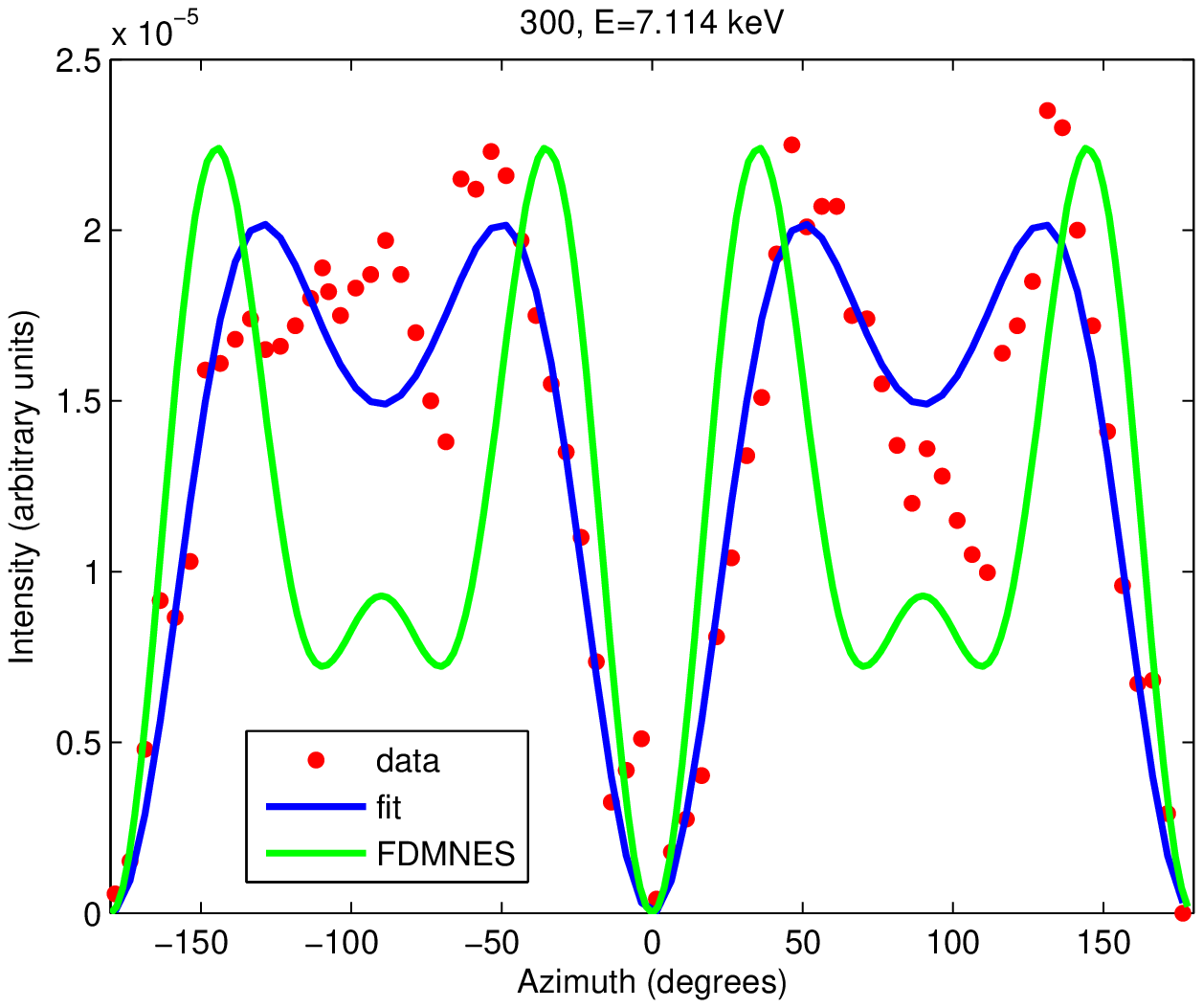}
    \includegraphics*[width=0.49\columnwidth]{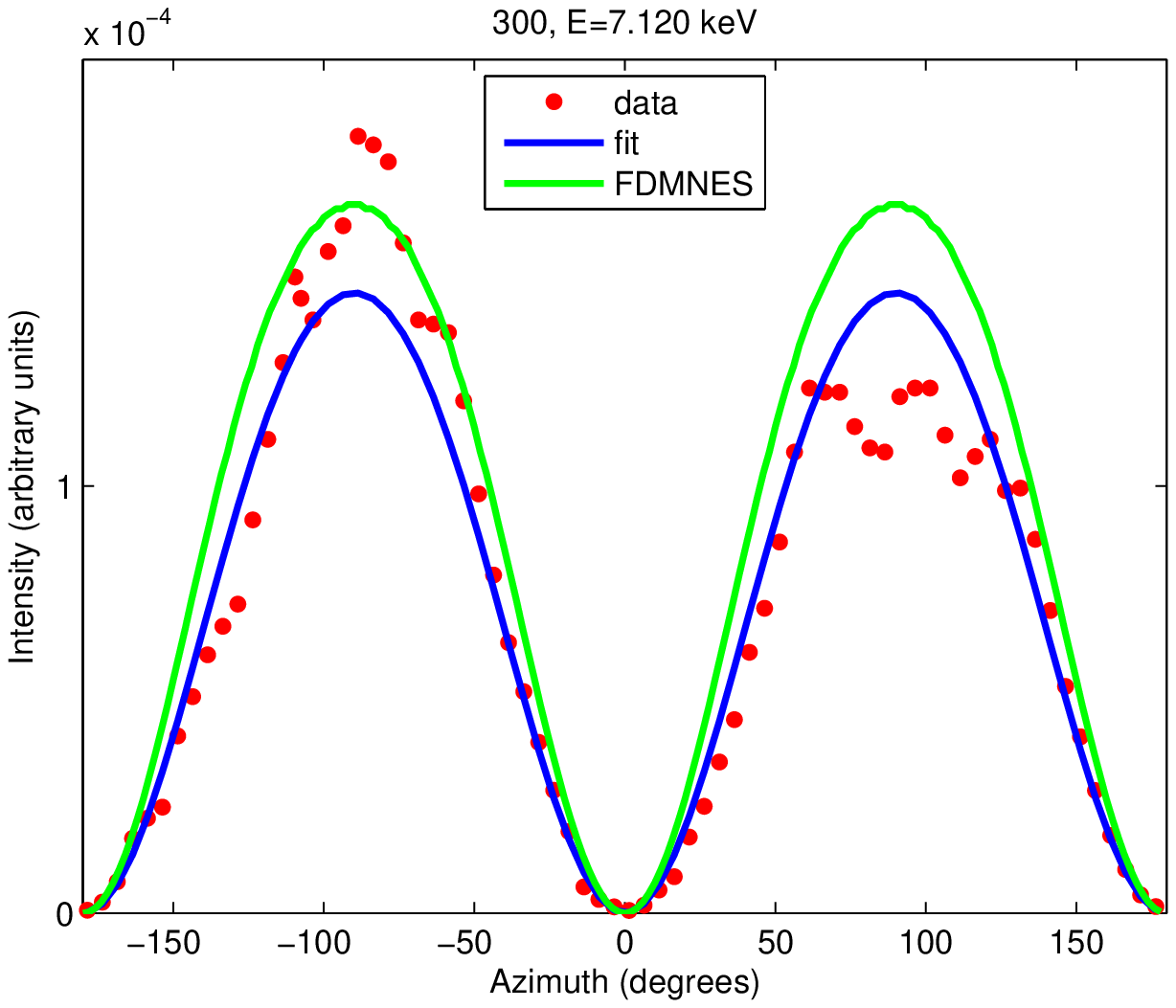}
    \caption{Azimuthal dependence of the reflection $300$ at 7.113 keV, 7.1135 keV, 7.114 keV and 7.12 keV. Comparison between experimental data ('data'), fit against equation (\ref{eq.rxsfit}) ('fit'), and simulation with FDMNES ('FDMNES'). The simulation fulfills equation \ref{eq.rxsfit}, but the coefficients are not the best fit to the data.}
    \label{fig.azimuth}
\end{figure}

Since  the main peaks, near and above the edge, result from the dipole-dipole scattering their intensities observed upon rotation about the scattering vector
 display a stable 2-fold azimuthal dependence.
In the pre-edge region, however,  the features observed  result  from the interplay between the dipole-quadrupole and quadrupole-quadrupole terms so that
more complex azimuthal dependencies are displayed which, in addition, vary strongly with energy (Fig. \ref{fig.azimuth}).
Their intensity values are  too small  to allow for detailed modelling,
but the FDMNES calculations using the previously determined convolution parameters
reproduce the general shape of the data reasonably well (Fig. \ref{fig.azimuth}).

\begin{figure}
    \includegraphics*[width=0.49\columnwidth]{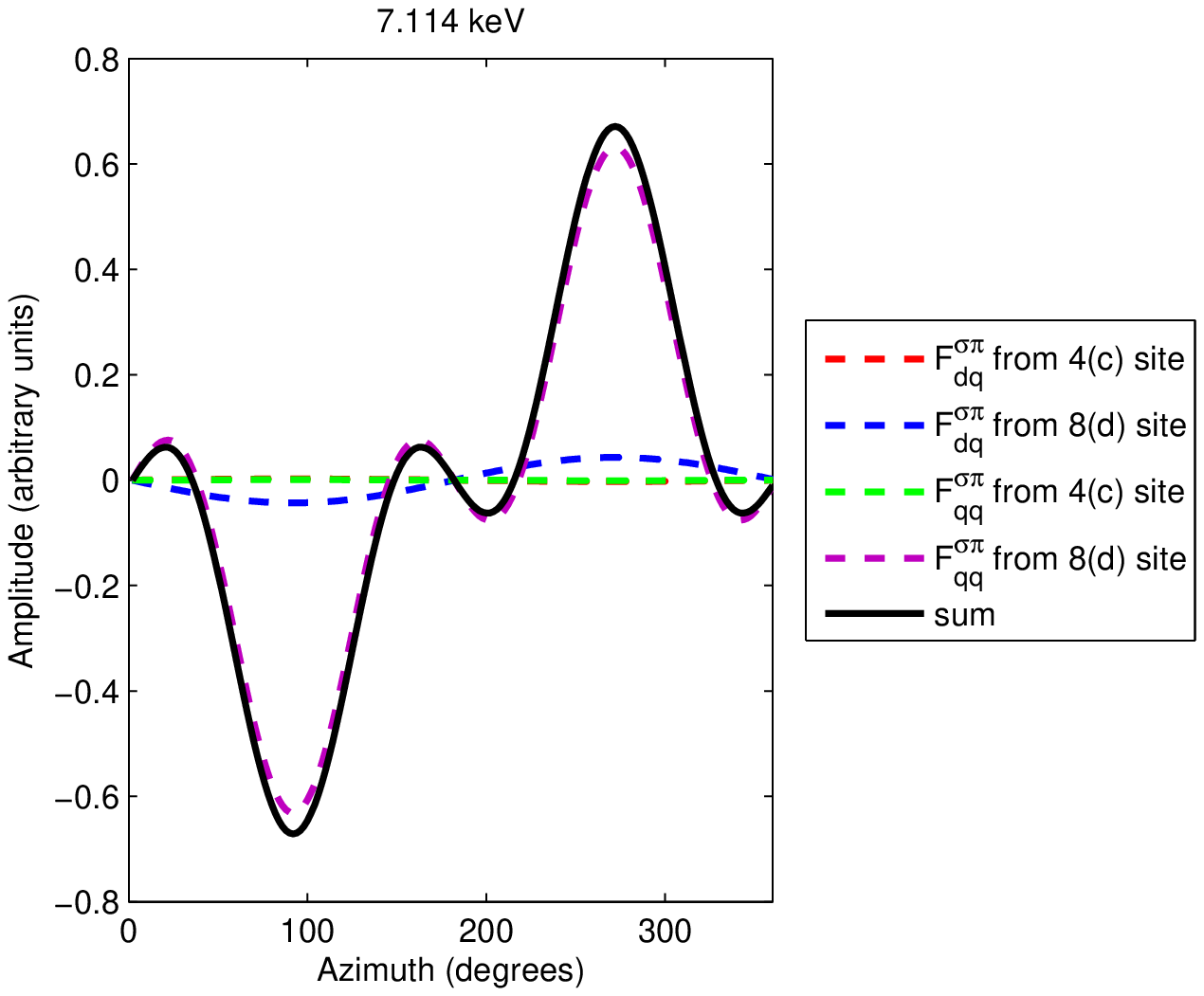}
    \includegraphics*[width=0.49\columnwidth]{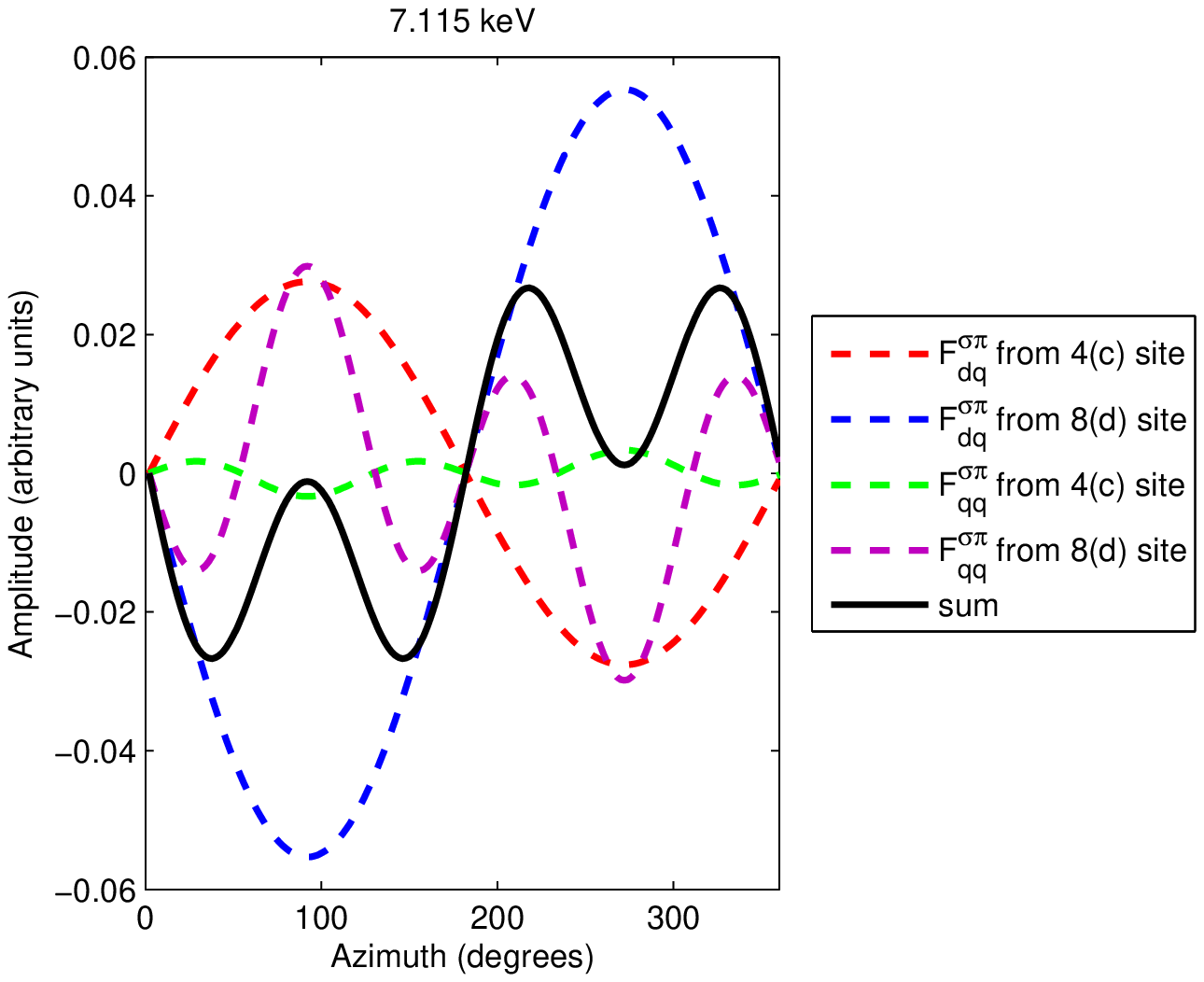}
    \caption{Azimuthal dependencies of the dipole-quadrupole and quadrupole-quadrupole contributions to the $300$ reflection at 7.114 keV (left) and 7.115 keV (right). The FDMNES convolution parameters are the same as previously determined.}
    \label{fig.contributions}
\end{figure}

Figure \ref{fig.contributions} shows the calculated dipole-quadrupole and the
quadrupole-quadrupole contributions to reflection 300 of each iron position,  at 7.114 keV and 7.115 keV.
Although these energies seem very close to each other, the various contributions show  strong energy dependent changes of azimuthal behaviour,
which affect the total intensity.

\section{Conclusions}

Resonant X-ray diffraction near the iron $K$-edge was studied
in iron orthoborate, both theoretically and experimentally.
The complex energy and azimuthal dependencies of the $300$, $500$ and $700$
forbidden reflection intensities are explained as resulting from the interference between the radiation
scattered from the two crystallographically inequivalent iron positions: $8(d)$ and $4(c)$.
In the dipole-dipole approximation, the $300$ intensity results mainly from Fe at the $8(d)$ position,
whereas the $500$ and $700$ intensities result respectively from constructive and destructive interferences of radiation scattered from Fe at both crystallographic positions.
The features in the pre-edge region can be explained by the interplay of dipole-quadrupole and quadrupole-quadrupole processes.
Due to the low point symmetry of the iron sites, the large number of parameters required for the full description of the resonant scattering prevented their complete determination.
Nevertheless, the dipole-dipole components could be extracted from the large part of the energy spectra where higher rank processes are comparatively weak, i.e. near and above the $K$-edge.
Theoretical calculations with the FDMNES code and subsequent fits of the experimental spectra suggest an energy difference of 0.7 $eV$ between the $K$-edges of the two non-equivalent iron atoms.
This chemical shift is attributed to the different crystallographic environments, to which RXS is very sensitive.
The measurements of the $700$ reflection, whose net intensity and energy spectrum were found stable over a wide temperature range,
indicate the absence of significant contributions of magnetic and/or thermal-motion-induced scattering.

The investigations demonstrate also the remarkable sensitivity of RXS to the interplay of various radiation contributions,
and simultaneously that a thorough analysis is possible despite its apparent non-triviality.
However, only studies of both energy and azimuthal dependencies of several forbidden reflection intensities
provide the possibility to distinguish contributions from different crystallographic sites and hence to study differences in the electronic interactions of atoms thereon.
Finally, it should be noted that excellent experimental data could be extracted
from measurements that were seriously affected by the poor crystal quality, frequent multiple scattering and strong self absorption.
This finding demonstrates as in earlier studies, e.g. \cite{kirfel91}, that RXS as a local probe is neither restricted to using nearly perfect crystals nor to especially favourable diffraction geometries.

\section{Acknowlegments}
The authors acknowledge the support of the XMaS staff for the RXS measurements.
This work was supported by grants RFBR 07-02-00324 and INTAS 01-0822.

\appendix
\setcounter{section}{1}
\section*{References}

\bibliographystyle{unsrt}
\bibliography{biblio_ortho}

\end{document}